\documentclass[lettersize,journal]{IEEEtran}

\usepackage{cite}
\usepackage{amsmath,amssymb,amsfonts}
\usepackage{graphicx}
\usepackage{textcomp}
\usepackage{xcolor}
\usepackage{multirow}
\usepackage{listings}
\usepackage{booktabs}
\usepackage{mathrsfs}
\usepackage{framed}
\usepackage[ruled, boxed,linesnumbered]{algorithm2e}
\usepackage{colortbl}
\usepackage{algpseudocode}
\usepackage{amsmath}
\usepackage{enumitem}
\usepackage{graphicx}
\usepackage{threeparttable}
\usepackage{float}
\usepackage{url}
\usepackage{graphicx}
\usepackage{subfigure}

\usepackage{bbding}
\usepackage{xcolor}

\usepackage{array}

\usepackage{wasysym}
\usepackage{pifont}
\usepackage{tcolorbox}
\usepackage[hidelinks]{hyperref}

\definecolor{verylightgray}{rgb}{.97,.97,.97}

\lstdefinelanguage{Solidity}{
	keywords=[1]{anonymous, assembly, assert, break, call, callcode, case, catch, class, constant, continue, constructor, contract, debugger, default, delegatecall, delete, do, else, emit, event, experimental, export, external, false, finally, for, function, gas, if, implements, import, in, indexed, instanceof, interface, internal, is, length, library, log0, log1, log2, log3, log4, memory, modifier, new, payable, pragma, private, protected, public, pure, push, require, return, returns, revert, selfdestruct, send, solidity, storage, struct, suicide, super, switch, then, this, throw, transfer, true, try, typeof, using, value, view, while, with, addmod, ecrecover, keccak256, mulmod, ripemd160, sha256, sha3}, 
	keywordstyle=[1]\color{blue}\bfseries,
	keywords=[2]{address, bool, byte, bytes, bytes1, bytes2, bytes3, bytes4, bytes5, bytes6, bytes7, bytes8, bytes9, bytes10, bytes11, bytes12, bytes13, bytes14, bytes15, bytes16, bytes17, bytes18, bytes19, bytes20, bytes21, bytes22, bytes23, bytes24, bytes25, bytes26, bytes27, bytes28, bytes29, bytes30, bytes31, bytes32, enum, int, int8, int16, int24, int32, int40, int48, int56, int64, int72, int80, int88, int96, int104, int112, int120, int128, int136, int144, int152, int160, int168, int176, int184, int192, int200, int208, int216, int224, int232, int240, int248, int256, mapping, string, uint, uint8, uint16, uint24, uint32, uint40, uint48, uint56, uint64, uint72, uint80, uint88, uint96, uint104, uint112, uint120, uint128, uint136, uint144, uint152, uint160, uint168, uint176, uint184, uint192, uint200, uint208, uint216, uint224, uint232, uint240, uint248, uint256, var, void, ether, finney, szabo, wei, days, hours, minutes, seconds, weeks, years},	
	keywordstyle=[2]\color{teal}\bfseries,
	keywords=[3]{block, blockhash, coinbase, difficulty, gaslimit, number, timestamp, msg, data, gas, sender, sig, value, now, tx, gasprice, origin},	
	keywordstyle=[3]\color{violet}\bfseries,
	identifierstyle=\color{black},
	sensitive=true,
	comment=[l]{//},
	morecomment=[s]{/*}{*/},
	commentstyle=\color{gray}\ttfamily,
	stringstyle=\color{red}\ttfamily,
	morestring=[b]',
	morestring=[b]"
}

\def\BibTeX{{\rm B\kern-.05em{\sc i\kern-.025em b}\kern-.08em
    T\kern-.1667em\lower.7ex\hbox{E}\kern-.125emX}}
\begin{document}

\author{Ruichao Liang, Jing Chen, Cong Wu, Kun He, Yueming Wu, Ruochen Cao, \\Ruiying Du, Yang Liu, Ziming Zhao\vspace{-15pt}
\thanks{Ruichao Liang, Jing Chen, Kun He, Ruochen Cao, and Ruiying Du are with Key Laboratory of Aerospace Information Security and Trusted Computing, Ministry of Education, School of Cyber Science and Engineering, Wuhan University, Wuhan, 430072, China. Email: {liangruichao, chenjing, hekun, crc2019, duraying}@whu.edu.cn. 
Cong Wu, Yang Liu, and Yueming Wu are with Cyber Security Lab, College of Computing and Data Science, Nanyang Technological University, Singapore. Email: {cong.wu, yangliu}@ntu.edu.sg, wuyueming21@gmail.com.
Ziming Zhao is with Department of Computer Science and Engineering, University at Buffalo, New York. Email: zimingzh@buffalo.edu.}
}

\title{\textsc{Vulseye}: Detect Smart Contract Vulnerabilities via Stateful Directed Graybox Fuzzing}

\maketitle

\begin{abstract}
	Smart contracts, the cornerstone of decentralized applications, have become increasingly prominent in revolutionizing the digital landscape.
	However, vulnerabilities in smart contracts pose great risks to user assets and undermine overall trust in decentralized systems.
  Fuzzing, a prominent security testing technique, is extensively explored to detect vulnerabilities.
	But current smart contract fuzzers fall short of expectations in testing efficiency for two primary reasons.
	Firstly, smart contracts are stateful programs, and existing approaches, primarily coverage-guided, lack effective feedback from the contract state. 
	Consequently, they struggle to effectively explore the contract state space.
	Secondly, coverage-guided fuzzers, aiming for comprehensive program coverage, may lead to a wastage of testing resources on benign code areas.
	This wastage worsens in smart contract testing, as the mix of code and state spaces further complicates comprehensive testing.

	To address these challenges, we propose \textsc{Vulseye}, a stateful directed graybox fuzzer for smart contracts guided by vulnerabilities.
	Different from prior works, \textsc{Vulseye} achieves stateful directed fuzzing by prioritizing testing resources to code areas and contract states that are more prone to vulnerabilities.
  We introduce \textit{Code Targets} and \textit{State Targets} into fuzzing loops as the testing targets of \textsc{Vulseye}.
  We use static analysis and pattern matching to pinpoint \textit{Code Targets}, and propose a scalable backward analysis algorithm to specify \textit{State Targets}.
	We design a novel fitness metric that leverages feedback from both the contract code space and state space, directing fuzzing toward these targets.
	With the guidance of code and state targets, \textsc{Vulseye} alleviates the wastage of testing resources on benign code areas and achieves effective stateful fuzzing.
	In comparison with state-of-the-art fuzzers, \textsc{Vulseye} demonstrated superior effectiveness and efficiency. 
	Notably, it uncovered 4,845 vulnerabilities in 42,738 real-world smart contracts, outperforming existing approaches by up to 9.7$\times$, and identified 11 previously unknown vulnerabilities within the top 50 Ethereum DApps, involving approximately 2,500,000 USD.

\end{abstract}


\section{Introduction}
\label{Introduction}
Smart contracts, self-executing programs that automate the requisite actions of agreements or contracts, form a cornerstone of the blockchain and cryptocurrency ecosystem.
However, vulnerabilities within these contracts pose considerable risks~\cite{bugs}, exemplified by the DAO incident, where a reentrancy vulnerability led to a loss of \$150 million~\cite{DAO}, and the Poly Network breach, resulting in a \$611 million loss~\cite{polynetwork}.
Thus, it is imperative to conduct thorough security testing on smart contracts to detect vulnerabilities and prevent significant financial and operational consequences.

Fuzz testing, an automated technique for program testing, identifies vulnerabilities by generating random inputs and systematically exploring a program's space~\cite{10.1145/3512345}.
Various efforts have been undertaken to apply fuzz testing techniques to smart contracts~\cite{10548588, ILF, ECHIDNA, Harvey, Targeted, sFuzz, ConFuzzius, Smartian, Effectively, Rethinking}. 
These initiatives involve generating inputs that mimic real-world transactions for testing smart contracts, with a focus on coverage-guided greybox fuzzing to improve code coverage and facilitate vulnerability discovery.
For example, \textsc{sFuzz}~\cite{sFuzz} introduces an adaptive strategy for seed selection aimed at reaching hard-to-cover branches for higher overall code coverage.
\textsc{Smartian}~\cite{Smartian} utilizes data-flow-based feedback for generating high-quality test cases in order to enhance the code coverage.
Despite these efforts, there still remain following significant gaps.

\textit{\textbf{\emph{i) Insufficient State Space Exploration in Smart Contract Fuzzing.}}}
Although existing contract fuzzers have thoroughly studied the use of feedback from code space, such as code coverage, to guide fuzzing, they lack an effective feedback mechanism for contract states, resulting in indiscriminate exploration of the contract state space.
As smart contracts are intrinsically stateful~\cite{SmarTest}, with state variables playing a pivotal role in their functionality, accurately identifying and examining specific states is essential for vulnerability exploitation.
Existing approaches such as \textsc{ConFuzzius}~\cite{ConFuzzius} and \textsc{Harvey}~\cite{Harvey} typically generate contract states through combinations of transactions.
\textsc{ITYFuzz}~\cite{ITYFuzz} utilizes snapshot for contract state to reduce the time required for transaction re-execution.
However, they are insufficient for a targeted and systematic examination of the contract state space, thereby impeding effective vulnerability detection.

\textit{\textbf{\emph{ii) Inefficient Code Space Exploration in Smart Contract Fuzzing.}}}
Most existing contract fuzzers adopt the coverage-guided fuzzing strategy, aiming for comprehensive program coverage.
However, research has shown that vulnerabilities within programs tend to be concentrated in specific code areas, leaving the majority of code benign~\cite{10.1145/3510003.3510230}.
Therefore, coverage-guided fuzzers may waste testing resources on these non-vulnerable areas~\cite{sok}.
This issue becomes more pronounced in smart contract testing, where the complex interaction between contract state and code spaces complicates comprehensive testing, resulting in a significant portion of testing efforts being allocated to benign code areas and contract states.

To bridge the gaps, we aim to design a stateful directed fuzzer for smart contracts, inspired by the concept of directed greybox fuzzing (DGF)~\cite{sok, 10.1145/3133956.3134020}.
DGF, known for its efficiency in allocating testing resources to specific areas of interest while minimizing unnecessary stress on unrelated parts, has shown effectiveness in various contexts~\cite{aflgo, Hawkeye, FISHFUZZ, SelectFuzz}.
However, existing DGFs lack capabilities for stateful fuzzing, primarily due to their focus on the code space while overlooking the essential state space in smart contracts. 
Moreover, these tools depend heavily on external information like bug reports for target identification~\cite{sok}, limiting their autonomy in identifying smart contract vulnerabilities. 
Different from the existing DGFs, our motivations include: i) achieving stateful fuzzing in the contract state space through the use of feedback from critical states, and ii) enabling directed fuzzing in the contract code space via automated target identification.

\textbf{Our Solution}. In this paper, we propose \textsc{Vulseye}, a stateful directed graybox fuzzer for smart contracts.
It prioritizes resources on vulnerable code areas and contract states, enhancing the efficiency of vulnerability detection.
We utilize static analysis and pattern matching to identify code areas vulnerable to exploits, termed as \textit{code targets}. 
We design a backward analysis algorithm for bytecode to determine the critical contract states required for vulnerability exploitation, termed as \textit{state targets}.
Code and state targets provide feedback in both code and state levels during the fuzzing process of \textsc{Vulseye}.
We propose a novel fitness metric integrating the feedback from code and state levels to achieve directed fuzzing across both contract code and state spaces.

We evaluated \textsc{Vulseye}'s performance on both closed datasets and real-world scenarios.
The evaluation reveals that \textsc{Vulseye} outperforms SOTA tools in vulnerability detection and proves effective in identifying real-world vulnerabilities.
Notably, \textsc{Vulseye} detected 4,845 vulnerabilities among 42,738 real-world smart contracts, outperforming SOTA tools by up to 9.7$\times$.
Besides, utilizing \textsc{Vulseye}, we found 11 previously unknown vulnerabilities in the top 50 Ethereum DApps and have reported them to corresponding authorities.

\textbf{Contributions.} This paper makes following contributions.
\begin{itemize}[leftmargin=4mm, itemindent=0mm]
  \item We propose \textsc{Vulseye}, the first stateful directed graybox fuzzer for smart contracts. 
  It autonomously identifies vulnerable code areas and contract states, then prioritizes testing resources on these targets, achieving stateful and directed exploration of both contract code and state spaces.

  \item We design a backward analysis algorithm for contract bytecode. 
  It efficiently identifies vulnerable states to specify state targets and provides feedback on the contract state space throughout the fuzzing process.
   
  \item We contribute a fitness metric that evaluates the proximity of seeds to testing targets at the contract code and state levels, effectively gauging the potential of seeds to trigger vulnerabilities.
  This algorithm guides seed scheduling in fuzzing, facilitating targeted vulnerability detection within both code and state spaces.

  \item We comprehensively evaluated \textsc{Vulseye} and benchmarked it against SOTA tools.
  The results show that: i) \textsc{Vulseye} reaches specific test targets up to 8.4$\times$ faster than the baselines.
  ii) It outperforms SOTA tools in terms of both efficiency and quantity on a ground truth vulnerability dataset.
  iii) It identified 4,845 vulnerabilities among 42,738 real-world smart contracts, surpassing SOTA tools in both quantity (9.7$\times$) and true positive rate.
  iv) Additionally, it uncovered 11 previously unknown vulnerabilities in the top 50 Ethereum DApps, involving about 2,500,000 USD.

\end{itemize}

\begin{table}[]
  \caption{Comparison of representative related works\tnote{*}.}
  \label{table 1}
  \setlength{\tabcolsep}{3.3pt}
  \scalebox{0.86}{
  \begin{threeparttable}
  \begin{tabular}{c|ccccc}
  \hline
  \textbf{Features}    & \textbf{\begin{tabular}[c]{@{}c@{}}\scalebox{0.8}{Data} \\ \scalebox{0.8}{Dependency}\tnote{\dag}\end{tabular}} & \textbf{\begin{tabular}[c]{@{}c@{}}\scalebox{0.8}{Directed Seed}\\ \scalebox{0.8}{Mutation}\end{tabular}} & \textbf{\begin{tabular}[c]{@{}c@{}}\scalebox{0.8}{Adaptive} \\ \scalebox{0.8}{Seed Priority}\end{tabular}} & \textbf{\begin{tabular}[c]{@{}c@{}}\scalebox{0.8}{Exploring} \\ \scalebox{0.8}{Code Space}\end{tabular}} & \textbf{\begin{tabular}[c]{@{}c@{}}\scalebox{0.8}{Exploring} \\ \scalebox{0.8}{State Space}\tnote{\ddag}\end{tabular}} \\ \hline
  \scalebox{0.85}{\textsc{ContractFuzzer}}~\cite{contractfuzzer}       & \Circle                                                                   & \Circle                                                                         & \Circle                                                                          & \LEFTcircle                                                                       & \Circle                                                                         \\
  \textsc{Harvey}~\cite{Harvey}               & \Circle                                                                   & \CIRCLE                                                                         & \Circle                                                                          & \CIRCLE                                                                        & \LEFTcircle                                                                        \\
  \textsc{sFuzz}~\cite{sFuzz}                & \Circle                                                                   & \LEFTcircle                                                                        & \Circle                                                                          & \CIRCLE                                                                        & \Circle                                                                         \\
  \textsc{ConFuzzius}~\cite{ConFuzzius}           & \CIRCLE                                                                   & \Circle                                                                         & \LEFTcircle                                                                         & \CIRCLE                                                                        & \LEFTcircle                                                                        \\
  \textsc{Smartian}~\cite{Smartian}             & \CIRCLE                                                                   & \Circle                                                                         & \Circle                                                                          & \CIRCLE                                                                        & \Circle                                                                         \\
  \textsc{IR-Fuzz}~\cite{Rethinking}              & \CIRCLE                                                                   & \Circle                                                                         & \CIRCLE                                                                          & \CIRCLE                                                                        & \Circle                                                                         \\
  \textsc{ITYFuzz}~\cite{ITYFuzz}              & \Circle                                                                   & \Circle                                                                         & \CIRCLE                                                                          & \CIRCLE                                                                        & \LEFTcircle                                                                         \\ \hline
  \textbf{\textsc{Vulseye}} & \CIRCLE                                                                   & \CIRCLE                                                                         & \CIRCLE                                                                          & \CIRCLE                                                                        & \CIRCLE                                                                         \\ \hline
  
\end{tabular}
\begin{tablenotes}
  \item[*]\CIRCLE = true, \LEFTcircle = partially true, \Circle = false.
  \item[\dag]Support data dependency analysis to generate meaningful transaction sequences.
  \item[\ddag]Support testing on specific contract states.
\end{tablenotes}
\end{threeparttable}}
\end{table}

\section{Background}

\subsection{Stateful Smart Contract}
Smart contracts~\cite{clack2016smart} are programs running on top of blockchain platforms.
They are Turing-complete programs used to automate the execution of an agreement and process assets on blockchain~\cite{Ethereum}.
Within smart contracts, there are a kind of variables called state variables that are created at the contract level and stored permanently on the blockchain~\cite{sailfish}.
The values of state variables can change based on the contract's execution, provide a direct reflection of the contract's states, and impact the contract's behavior and functionality.
State variables make smart contracts \textit{stateful} programs by preserving and managing their states throughout their lifecycle.

\subsection{Graybox Fuzzing}
Greybox fuzzing is a software testing method that combines aspects of white-box~\cite{godefroid2008automated} and black-box testing~\cite{DBLP:conf/uss/KimKBBKT23}.
It involves injecting semi-random input data, known as seeds, into a program to uncover vulnerabilities and potential issues~\cite{8371326}.
Based on the feedback information from the execution of the program under test cases, greybox fuzzers generate new inputs and perform seed scheduling to manage and prioritize these inputs to effectively test the program.
Most greybox fuzzing tools are coverage-guided, striving to explore as many program paths as possible~\cite{sok}.

\subsection{Limitations of Existing Contract Fuzzers}
Firstly, being stateful programs, smart contracts differ from typical programs by not only having their code space but also having an implicit state space.
This complexity adds a layer of difficulty to the exploration of the program space to detect vulnerabilities~\cite{281370}.
Some existing fuzzers take into account the stateful nature of smart contracts. 
For instance, some~\cite{ConFuzzius,Harvey,Smartian,Rethinking} attempt to manipulate the contract state by leveraging the read-write relationship of variables to create transaction sequences, while some~\cite{ITYFuzz} employ snapshot to preserve the contract state for rapid resumption. 
However, these approaches lack knowledge of the specific state required to trigger vulnerabilities, resulting in blind and inefficient exploration of the contract state space, as shown in Table~\ref{table 1}.

Secondly, existing tools predominantly rely on coverage-guided fuzzing strategies to achieve greater code coverage. 
However, this may result in an inefficient allocation of effort on testing non-vulnerable code areas, while overlooking critical code areas.
The stateful nature of smart contracts further compounds this issue.

These limitations are exemplified in Section~\ref{Motivating Example}. 
As shown in Table~\ref{table 1}, \textsc{Vulseye} surpasses existing state-of-the-art (SOTA) fuzzers by facilitating targeted exploration in both the code and state spaces of smart contracts.
These advancements collectively lead to more efficient vulnerability detection.

\begin{figure}[t!]
\centering
\begin{minipage}{0.48\textwidth}
\begin{lstlisting}[language = Solidity]

contract FancyBank{
  mapping(address => uint256) private balances;
  uint256 dueDate = 0, unlock = 0;
  event WithdrawalFailed(address user, uint256 amount);

  function deposit(uint256 amount) public payable{
    require(msg.value >= amount)
    balances[msg.sender] += amount;}

  function setState(uint256 time,uint256 State) public{
    dueDate = time;
    unlock = State;}

  function withdraw(uint256 amount) public {
    require(balance[msg.sender] >= amount);
    if(dueDate >30 && dueDate < 40 && unlock == 1){
      msg.sender.call{value: amount}();
      balance[msg.sender] -= amount;
    }else{
      emit WithdrawalFailed(msg.sender, amount);}}
  /** Other functions */
}

\end{lstlisting}
\end{minipage}
\caption{Motivating Example}
\label{figure 1}
\end{figure}

\section{Motivating Example}
\label{Motivating Example}
This section presents a motivating example: a contract that allows users to deposit funds and withdraw them as needed.
The contract's simplified segment is illustrated in Figure~\ref{figure 1}. 
It is vulnerable to reentrancy because of the \texttt{withdraw()} function.
Intended for users to withdraw funds, this function mistakenly transfers funds before updating the \texttt{balance} variable. 
This allows an attacker to reenter the \texttt{withdraw()} function multiple times during the external call in Line 19, ultimately draining the contract's funds.
However, it is not easy to exploit this vulnerability in practice, as illustrated below.

\subsection{State Feedback is Necessary}
As discussed in Section~\ref{Introduction}, stateful programs typically necessitate reaching specific states before the vulnerability is manifested.
In the contract depicted in Figure~\ref{figure 1}, invoking the \texttt{setState()} function to set the contract states to: \texttt{dueDate$\in$(30,40)} and \texttt{unlock$\in$\{1\}} is necessary before triggering the reentrancy vulnerability in Line 19.

Now considering a scenario with three seeds, each seed comprises two transactions triggering two functions:
\begin{itemize}[]
  \item[] \textit{Seed A}:  \ding{172}\texttt{deposit(10)} $\longrightarrow$  \ding{173}\texttt{withdraw(10)}
  \item[] \textit{Seed B}:  \ding{172}\texttt{setState(100,0)} $\longrightarrow$  \ding{173}\texttt{withdraw(10)}
  \item[] \textit{Seed C}:  \ding{172}\texttt{setState(50,1)} $\longrightarrow$  \ding{173}\texttt{withdraw(10)}
\end{itemize}
While all three seeds ultimately follow the \textit{else} branch (Line 21) and miss the vulnerable code location (Line 19), seed C is closer to exploiting the vulnerability.
This is because seed C modifies the critical state variables to be closer to the states that are necessary for triggering the vulnerability (i.e., \texttt{dueDate$\in$(30,40), unlock$\in$\{1\}}).
Consequently, allocating more testing resources to seed C during the fuzzing process can expedite the discovery of the vulnerability.
However, current approaches such as \textsc{ConFuzzius}~\cite{ConFuzzius} and \textsc{Harvey}~\cite{Harvey} rely solely on combining transactions in a Read-After-Write order (a trait present in all three seeds in this example) to manipulate the contract state, but they lack feedback from the contract state for systematic stateful exploration.
As a result, they fail to recognize \texttt{dueDate} and \texttt{unlock} as key state variables for triggering the vulnerability, thus hindering their ability to differentiate between the relative effectiveness of these three seeds.
This observation underscores the significance of using feedback from the state space in the fuzzing of stateful programs like smart contracts.

\subsection{Pursuing Comprehensive Coverage May Reduce Efficiency}

The vulnerability can be exploited when an attacker manages to pass the \textit{if} condition in Line 18, and execute the external call in Line 19.
However, exploiting this vulnerability in the \textit{if} branch is time-consuming due to the strict state conditions it requires. 
Test cases are more likely to take the \textit{else} branch.
In such situations, the current approaches with a coverage-guided fuzzing strategy would typically shift focus to other non-critical functions to enhance overall code coverage, since intensifying code coverage through testing the \texttt{withdraw()} function has presented significant challenges.
This observation underscores the inefficiency in blindly chasing code coverage, as it often results in disproportionate testing resources being spent on non-vulnerable code sections. 
A more effective strategy for uncovering vulnerabilities would be to strategically focus testing efforts on code segments with a higher susceptibility to vulnerabilities.

\section{Overview}
\label{overview}

We brief threat model and give an overview of \textsc{Vulseye}.

\begin{figure*}
	\centering
	\includegraphics[width=6in]{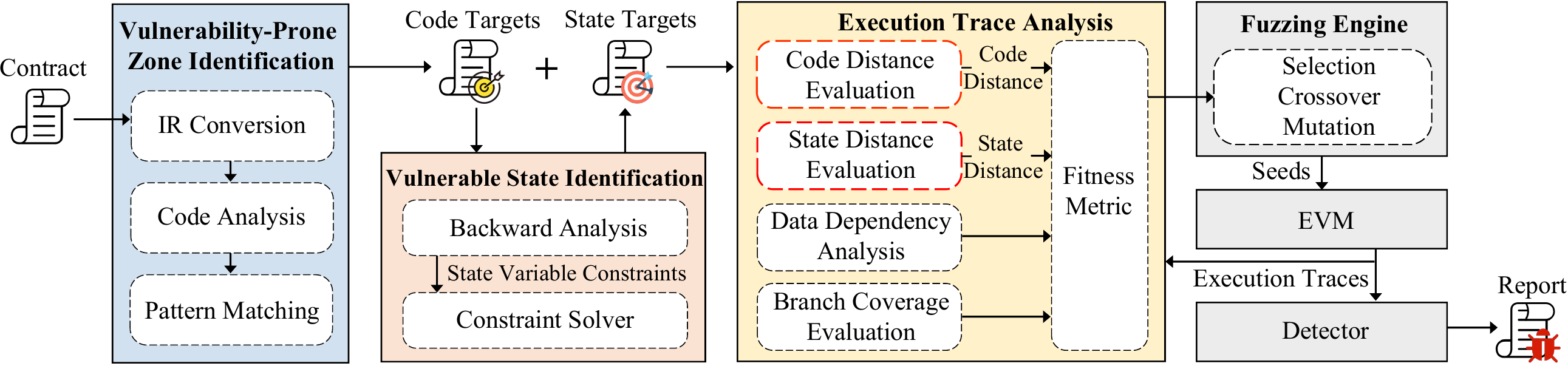}
	\caption{Overview of \textsc{Vulseye}.}
	\label{figure 2}
\end{figure*}
\vspace{-10pt}

\subsection{Threat Model}
Smart contracts, with their nature of blockchain-based execution and capability of handling digital assets, introduce a set of unique vulnerabilities distinct from traditional platforms.
These vulnerability threats often arise from developer oversights or inherent blockchain characteristics, and exploiting them can cause significant and irreversible financial damage.
They can be categorized into different types including \textit{Ether Leaking}, \textit{Reentrancy}, \textit{Controlled Delegatecall}, \textit{Suicidal}, etc., which have been well illustrated in previous studies~\cite{contractfuzzer,Effectively}.
The attacker has the ability to access data on the public blockchain, analyze smart contracts to identify vulnerabilities, and knows how to exploit them by sending transactions or deploying new contracts.

\textsc{Vulseye} aims to provide an efficient stateful directed fuzzing solution for vulnerability detection, ensuring the safety of smart contracts before deployment and enabling defenders to timely implement preventive measures for on-chain vulnerable contracts.
This will mitigate the threats that contract vulnerabilities pose to blockchain and decentralized ecosystems.

\subsection{Overview of Vulseye}
To implement stateful directed fuzzing in both contract code and state spaces to hunt vulnerabilities, we address three main challenges: \textbf{\textit{locating vulnerability-prone code}} in smart contracts, \textbf{\textit{identifying critical contract states}} required to exploit vulnerabilities in these code areas, and \textbf{\textit{guiding fuzzing to test specific states and code}} for accelerated vulnerability detection.

\begin{algorithm}[t!]\footnotesize
	\SetAlgoLined
	\SetAlgoNlRelativeSize{-1}
	\SetKwInOut{Input}{Input}
	\SetKwInOut{Output}{Output}
	\caption{Code Target Identification.}
	\label{algorithm 0}
	\Input{\texttt{Contract} $\mathcal{C}$}
	\Output{\texttt{CodeTargets}}
	\SetNlSty{}{}{:}
	\SetNlSkip{0.5em}
	\SetNlSkip{-1.2em}
	\SetNlSkip{0.5em}
	\texttt{irCode}, \texttt{cfg} $\gets$ \textit{IRConversion} ($\mathcal{C}$)\\
	\texttt{res} $\gets$ \textit{CodeAnalysis}(\texttt{irCode}, \texttt{cfg})\\
	\For{\texttt{node} in \texttt{cfg.nodes}}{
	  \For{\texttt{ir} in \texttt{node.irs}}{
		\texttt{offset} $\gets$ \textit{PatternMatching} (\texttt{ir,cfg,res})\\
		\texttt{CodeTargets} $\gets$ \textit{Locating} (\texttt{offset,cfg})
	}
	}
  \end{algorithm}
\setlength{\textfloatsep}{5pt}

As shown in Figure~\ref{figure 2}, to locate vulnerable code areas, a smart contract is first compiled and converted into an Intermediate Representation (IR), which serves as the foundation for subsequent static code analysis.
Based on the result of code analysis, \textsc{Vulseye} conducts pattern matching to identify code areas that could potentially harbor vulnerabilities, referred to as code targets.
To identify vulnerable contract states necessary for exploiting vulnerabilities, \textsc{Vulseye} conducts backward analysis on the control flow graph (CFG) starting from code targets, gathering the constraints associated with state variables.
Then, a constraint solver is applied to resolve the state constraints and specify the state targets.
During the fuzzing loop, \textsc{Vulseye} analyzes execution traces, evaluating both the code and state distances of seeds to estimate their propensity to trigger vulnerabilities. 
Seeds with a greater potential to trigger vulnerabilities are given higher priority in the seed scheduling process, and values from state targets are incorporated into the mutation pool to generate vulnerable contract states.
Vulnerability detection is carried out by a detector that evaluates the execution trace based on test oracles.

\begin{table}[t]
  \caption{Hazardous behaviors that may lead to vulnerabilities.}
  \label{table 2}
  \centering
  \begin{tabular}{lc}
  \hline
  \multicolumn{1}{c|}{\textbf{Hazardous Behaviors Description}}                                                                                                                                               & \textbf{Potential Bugs}                                            \\ \hline
  \multicolumn{1}{l|}{\begin{tabular}[c]{@{}l@{}}Has \textcolor{violet}{\textit{Payable}} function.\\ No \textcolor{blue}{\textit{Suicide}} or \textcolor{blue}{\textit{Selfdestruct}} operation.\\ No high-level or low-level \textcolor{blue}{\textit{calls}} that send ether.\end{tabular}}                             & Lock Ether                                                         \\ \hline
  \multicolumn{1}{l|}{\textcolor{blue}{\textit{Delegatecall}} using \textcolor{violet}{\textit{msg.data}} as destination.}                                                                                                                                           & \begin{tabular}[c]{@{}c@{}}Controlled \\ Delegatecall\end{tabular} \\ \hline
  \multicolumn{1}{l|}{\textcolor{blue}{\textit{Delegatecall}} using \textcolor{violet}{\textit{msg.data}} as arguments.}                                                                                                                                             & \begin{tabular}[c]{@{}c@{}}Dangerous \\ Delegatecall\end{tabular}  \\ \hline
  \multicolumn{1}{l|}{\begin{tabular}[c]{@{}l@{}}Use \textcolor{violet}{\textit{block data}} in \textcolor{violet}{\textit{Require}} or \textcolor{violet}{\textit{If}} statement.\\ Send ether following this statement.\end{tabular}}                                                              & \begin{tabular}[c]{@{}c@{}}Block \\ Dependency\end{tabular}        \\ \hline
  \multicolumn{1}{l|}{\begin{tabular}[c]{@{}l@{}}\textcolor{blue}{\textit{Read}} a state variable.\\ External \textcolor{blue}{\textit{call}} after the \textcolor{blue}{\textit{read}}.\\ \textcolor{blue}{\textit{Write}} the same state variable after the \textcolor{blue}{\textit{call}}.\end{tabular}}                                        & Reentrancy                                                         \\ \hline
  \multicolumn{1}{l|}{\begin{tabular}[c]{@{}l@{}}No use of \textcolor{violet}{\textit{msg.sender}} as index.\\ Send ether with no dependency on \textcolor{violet}{\textit{msg.value}}.\\ Use \textcolor{violet}{\textit{msg.sender}} or \textcolor{violet}{\textit{msg.data}} as receiver.\end{tabular}}                        & \begin{tabular}[c]{@{}c@{}}Arbitrary \\ Send Ether\end{tabular}    \\ \hline
  \multicolumn{1}{l|}{\begin{tabular}[c]{@{}l@{}}Has function that is not protected.\\ Set this function as \textcolor{violet}{\textit{public}} or \textcolor{violet}{\textit{external}} visible.\\ Has \textcolor{blue}{\textit{Suicide}} or \textcolor{blue}{\textit{Sefldestruct}} in this function.\end{tabular}} & Suicidal                                                           \\ \hline
  \end{tabular}

  \end{table}

\section{Methodology}
\label{Vulseye}
In this section, we detail the design of \textsc{Vulseye}.

\subsection{Target Identification}
To implement directed gray box fuzzing, the first step is to specify potential vulnerable locations as test targets, which, in our solution, involves identifying code and state targets. 

\textbf{Vulnerability-prone Zone Identification.}
We statically analyze the contracts to identify potentially vulnerable code areas prior to conducting fuzzing, as illustrated in Algorithm~\ref{algorithm 0}.
We utilize SlithIR~\cite{feist2019slither}, an intermediate representation, to represent Solidity code of the contracts under test.
A series of code analyses are carried out based on the intermediate representation, such as function property analysis and data dependency analysis (Line 2).
Utilizing insights gained from these analyses, we navigate through the contract's CFG, scrutinizing semantic contract behaviors through pattern matching to identify hazardous behaviors that could potentially lead to vulnerabilities (Lines 3-5).
The rules of pattern matching are tailored to specific vulnerable behaviors, as outlined in Table~\ref{table 2}.
For instance, a sequence of basic blocks where a state variable is \ding{172} initially read before an external call, and \ding{173} subsequently written after the external call, may suggest a potential \textit{Reentrancy} vulnerability.
To enhance the detection of vulnerable code areas and reduce the risk of false negatives, we traverse the contract code space and conduct pattern matching with an over-approximation principle.
To pinpoint these hazardous behaviors, we locate them within basic blocks of the contract's CFG as the \textit{code targets} (Line 6).

It is important to note that, the \textit{code targets} identified in this phase do not directly report vulnerabilities; rather, they highlight potential vulnerable code areas (which may not necessarily result in vulnerabilities).
These flagged areas undergo further validation through subsequent fuzzing tests, thereby mitigating the risk of high false positive rates inherent in static analysis~\cite{8967437}.

\textbf{Vulnerable State Identification.}
Smart contracts are stateful programs, and vulnerabilities necessitate specific states for exploitation.
While we've identified vulnerable code targets and plan to utilize them as guidance for fuzzing, blindly altering the contract's state can still be inefficient in reaching those code targets and exploiting vulnerabilities.
To address this challenge, we introduce a backward analysis algorithm designed to identify vulnerable contract states, facilitating the stateful exploration in the contract state space.
Unlike symbolic execution~\cite{10.1145/2976749.2978309,10.1145/3274694.3274737,DBLP:conf/acsac/NikolicKSSH18} which is slow and suffers from poor scalability due to the path explosion problem, we focus solely on the code targets-related paths and states, ensuring the efficiency and scalability of our tool.
Algorithm~\ref{algorithm 1} reveals how we perform backward analysis on the contract's CFG and define the corresponding state targets.
Algorithm~\ref{algorithm 2} further details the process by which backward analysis systematically gathers all state constraints for a given path.

\setlength{\textfloatsep}{16pt}
\begin{algorithm}[t!]\footnotesize
  \SetAlgoLined
  \SetAlgoNlRelativeSize{-1}
  \SetKwInOut{Input}{Input}
  \SetKwInOut{Output}{Output}
  \caption{State Target Identification.}
  \label{algorithm 1}
  \Input{\texttt{CFG} $\mathcal{C}$, \texttt{Target} $\mathcal{T}$}
  \Output{\texttt{stateTarget}}
  \SetNlSty{}{}{:}
  \SetNlSkip{0.5em}
  \SetNlSkip{-1.2em}
  \SetNlSkip{0.5em}
  \texttt{stateCons} $\gets$ $\emptyset$\\
  \texttt{paths} $\gets$ \textit{FindPath} ($\mathcal{C}$, $\mathcal{T}$)\\
  \For{\texttt{p} in \texttt{paths}}{
  \texttt{pathCons} $\gets$ $\mathbb{U}$\\
  \texttt{branchPoints} $\gets$ \texttt{p.branchPoints}\\
  \For{\texttt{bp} in \texttt{branchPoints}}{
    \texttt{branchCons} $\gets$ \textit{BackwardAnalysis} (\texttt{p}, \texttt{bp})\\
    \texttt{pathCons} $\gets$ \texttt{pathCons} $\cap$ \texttt{branchCons}
  }
  \texttt{stateCons} $\gets$ \texttt{stateCons} $\cup$ \texttt{pathCons}
  }
  \texttt{stateTarget} $\gets$ \textit{Solve} (\texttt{stateCons})

\end{algorithm}

\setlength{\textfloatsep}{1pt}

\begin{algorithm}[t!]\footnotesize
  \SetAlgoLined
  \SetAlgoNlRelativeSize{-1}
  \SetKwInOut{Input}{Input}
  \SetKwInOut{Output}{Output}
  \caption{Contract Bytecode Backward Analysis.}
  \label{algorithm 2}
  \Input{\texttt{path} $\mathcal{P}$, \texttt{branchPoint} $\mathcal{B}$}
  \Output{\texttt{branchCons}}
  \SetNlSty{}{}{:}
  \SetNlSkip{0.5em}
  \SetNlSkip{-1.2em}
  \SetNlSkip{0.5em}
  \texttt{branchCons} $\gets$ \textit{InitCons ($\mathcal{P}$, $\mathcal{B}$)}\\
  \texttt{t} $\gets$ \textit{InitTraceBack ($\mathcal{P}$, $\mathcal{B}$)}\\
  \texttt{stack} $\gets$ []\\
  \texttt{reversedIns} $\gets$ $\mathcal{P}$\texttt{.BackwardFrom($\mathcal{B}$)}\\
  \While{\textit{NotEmpty (\texttt{t})} \&\& \textit{NotEmpty (\texttt{reversedIns})}}{
  \texttt{ins}, \texttt{pc} $\gets$ \texttt{reversedIns.Pop()}\\
  \texttt{stack[pc]} $\gets$ \textit{StackReconstruction (\texttt{ins}, \texttt{pc}, \texttt{stack})}\\
  \texttt{cons}, \texttt{t} $\gets$ \textit{ConsCollection (\texttt{ins}, \texttt{pc}, \texttt{stack}, \texttt{t})}\\
  \texttt{branchCons} $\gets$ \texttt{branchCons} $\cap$ \texttt{cons}\\

  }
\end{algorithm}
\setlength{\textfloatsep}{10pt}

Algorithm~\ref{algorithm 1} takes as input the contract's CFG and the code target obtained from the vulnerability-prone zone identification process, and outputs a set of state targets.
We handle loops in the CFG using loop unrolling with a upper bound of 20.
This compromise balances efficiency and soundness in our analysis, as smart contracts typically avoid extensive loops due to gas limits.
We initialize the set of state constraints (\texttt{stateCons}) and use NetworkX~\cite{networkx}, an efficient tool for manipulating graph structures, to identify all paths leading to the target basic blocks (Lines 1-2).
For each path, we extract the involved branch jumps (typically the \texttt{JUMPI} instruction) in Line 5 and conduct backward analysis at these branch points, gathering constraints on state variables in Line 7.
Next, we intersect these constraints in Line 8 to derive state requirements for this path to reach the code target.
The union of these path-specific constraints (Line 10) defines the overall state requirements for all paths.
Finally, we employ a constraint solver to resolve these constraints, determining the specific contract states required to reach the code target, referred to as the \textit{state targets}.
It is worth noting that for state variables of mapping type that take a symbolic address as the key, we do not regard them as state targets because the key cannot be concretely determined in static analysis.

Algorithm~\ref{algorithm 2} further details the implementation of backward analysis at each branch point.
In Line 1, we generate an initial constraint based on the jump direction at the current branch point.
To be specific, the jump direction is determined by the second parameter of the \texttt{JUMPI} instruction.
If the second parameter is non-zero, the program jumps to the \textit{if} branch; otherwise, it proceeds to the \textit{else} branch.
As the concrete values of the instruction parameters cannot be known statically, we use symbolic variables to represent them.
To transform this initial constraint into a state-related constraint, we designate the symbolic variables in this constraint as the trace-back target (Line 2) and collect their relationships with state variables during the backward traversal of instructions (Lines 5-10).
In Line 7, we reconstruct the EVM stack by analyzing instructions in reverse order.
Utilizing the recovered stack information, we generate new constraints derived from the initial constraint, subsequently refreshing the trace-back targets in a last-in-first-out (LIFO) manner in Line 8.
This iterative process continues until all traced targets manifest as either state variables or state-independent constants, yielding a set of state-related constraints.

\begin{figure}[t!]
	\centering
	\scalebox{0.9}{
	\includegraphics[width=2.7in]{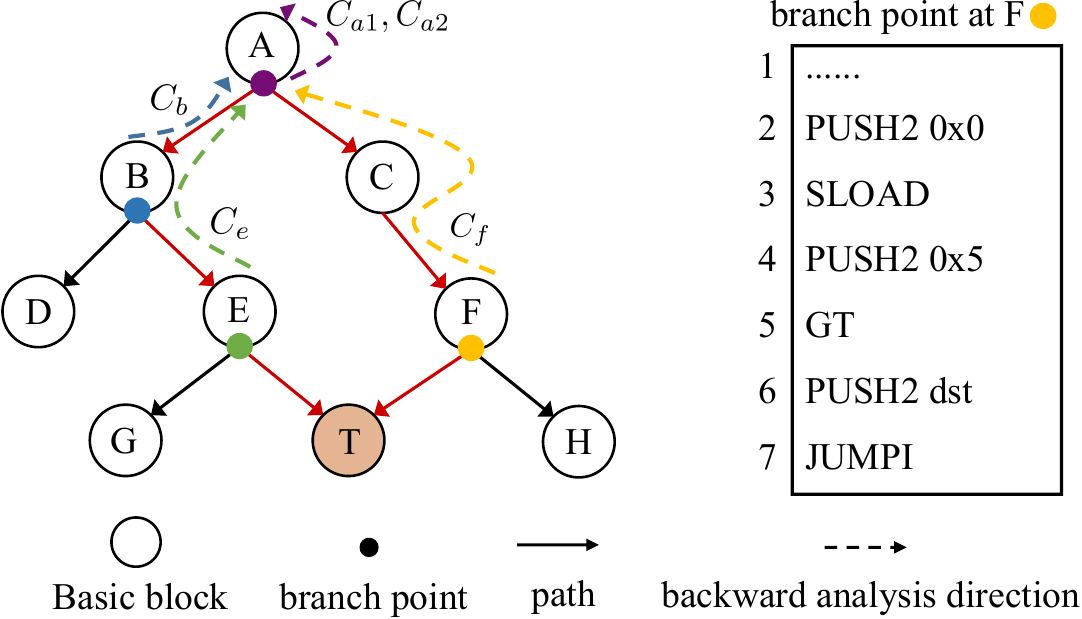}}
	\caption{Example for State Identification.}
	\label{figure 3}
\end{figure}

  \begin{figure*}[t]
	\centering
  
	\includegraphics[width=6.4in]{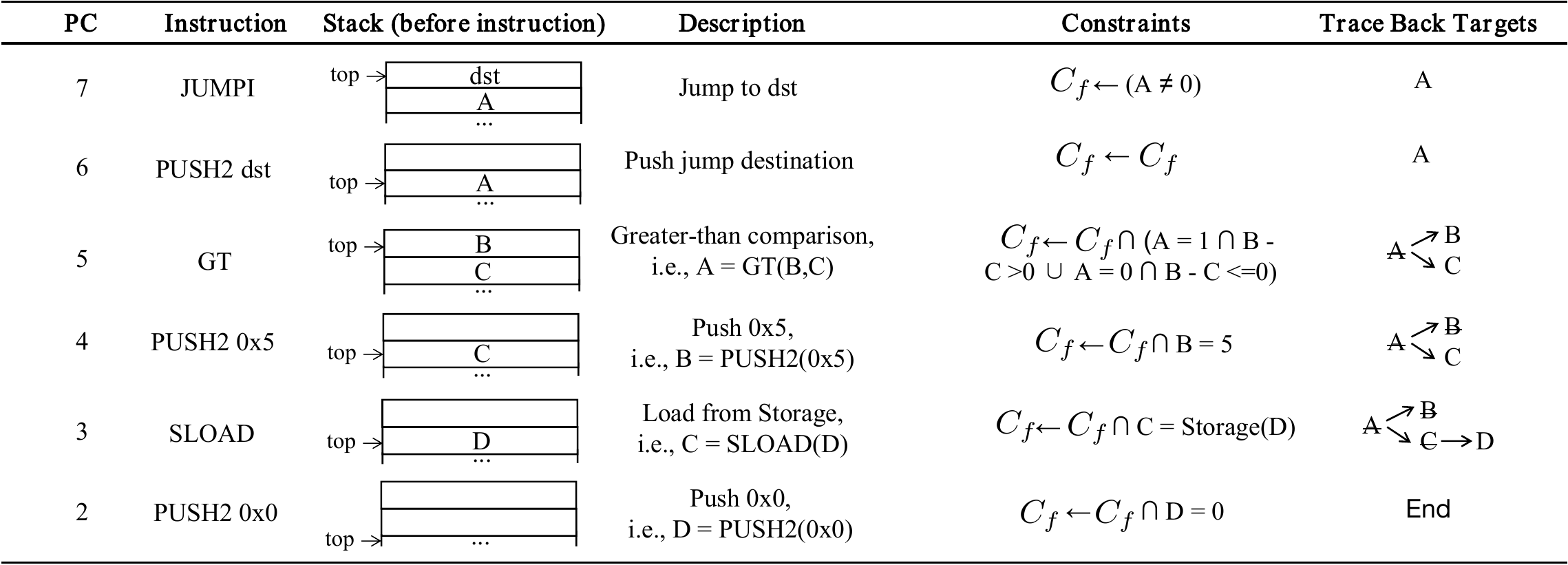}
	\caption{Example for Backward Analysis at the branch point F.}
	\label{figure 4}
  \end{figure*}

\textbf{Example for State Identification.}
In Figure~\ref{figure 3}, the left side illustrates a simplified contract CFG with a determined code target, denoted as node T.
As described in Algorithm~\ref{algorithm 1}, there are two paths reaching the target, i.e., A-B-E-T and A-C-F-T, with branch points at nodes A, B, E, and F.
Backward analysis on these branch points yields constraints $C_{a1}$, $C_{a2}$, $C_{b}$, $C_{e}$, and $C_{f}$. 
The state constraints for the code target are then expressed as $(C_{a1} \cap C_{b} \cap C_{e}) \cup (C_{a2} \cap C_{f})$.
The right side of Figure~\ref{figure 3} is a simplified opcode segment of the branch point at node F, and Figure~\ref{figure 4} demonstrates the backward analysis process at this branch point.
In Figure~\ref{figure 4}, columns one and two display the program counter and corresponding instructions, while the third column represents the EVM stack information restored based on these instructions.
The fourth column describes the current instruction's impact on stack elements, and the fifth column presents constraints generated during backward analysis. 
The sixth column displays elements in the trace-back list.
Backward analysis initiates at the \texttt{JUMPI} instruction (pc=7), where the top of the EVM stack contains two elements, i.e., the jump destination \texttt{dst} and value \texttt{A}.
The \texttt{JUMPI} instruction specifies that if $\texttt{A} \neq 0$, pc jumps to dst; otherwise, it jumps to pc+1. 
Consequently, the first constraint, $\texttt{A} \neq 0$, is derived, based on the jump direction, and the value \texttt{A} is added to the trace-back list.
The \texttt{PUSH2} instruction (pc=6) pushes \texttt{dst} onto the stack, which does not affect the tracing target.
Analyzing the \texttt{GT} instruction (pc=5) which compares the sizes of \texttt{B} and \texttt{C} and then assigns the result to \texttt{A}, produces a new constraint: $(\texttt{A}=1 \cap \texttt{B}-\texttt{C}>0) \cup (\texttt{A}=0 \cap \texttt{B}-\texttt{C}<=0)$.
Consequently, the element in the trace-back list transitions from \texttt{A} to \texttt{B} and \texttt{C}.
The subsequent \texttt{PUSH2} instruction (pc=4) reveals \texttt{B} as 5, and the \texttt{SLOAD} instruction (pc=3) indicates \texttt{C} is obtained from the current contract states at slot \texttt{D}. 
The backward analysis continues until the trace-back list is empty.
These derived constraints constitute the state constraints for this branch point, denoted as $C_{f}$ in Figure~\ref{figure 3}

\subsection{Execution Trace Analysis}
We analyze the execution trace of each seed during the fuzzing loop and employ a fitness metric algorithm to assess the proximity of the seeds to both code and state targets. 
This evaluation serves to measure the likelihood of a seed triggering a vulnerability. 
By utilizing this approach, we can strategically allocate testing resources to prioritize better seeds, thereby accelerating the testing process.

\textbf{Code Distance Evaluation.}
The code target is essentially a program basic block in the contract's CFG.
In this section, we refer to it as the target block.
For a basic block $BB$, we define its distance to a target block $TB$ as
\begin{equation}\footnotesize
  \label{eq1}
  d_{TB}(BB)=\left\{
  \begin{aligned}
  &d(BB,TB) \quad if \  BB \rightsquigarrow TB, \\
  &\infty \quad \quad \ \quad  \ \quad Otherwise.
  \end{aligned}
  \right.
\end{equation}
where $BB \rightsquigarrow TB$ represents that $TB$ is reachable from $BB$, and $d(BB,TB)$ represents the number of edges covered by the shortest path from $BB$ to $TB$.
Based on the above definition, for the block $BB$, we define its block distance $D(BB)$, as the harmonic mean of its distances to all target blocks, denoted as
\begin{equation}\footnotesize
  \label{eq2}
  D(BB)=\left\{
  \begin{aligned}
  &0\quad \quad \quad \quad \quad if \  BB \in \mathcal{Q}, \\
  &undefined\ \quad if \ (TB \in \mathcal{Q} \ | \  d_{TB}(BB) \equiv \infty),\\
  &\left| \mathcal{Q} \right|\times[\sum_{TB \in \mathcal{Q}} d_{TB}(BB)^{-1}]^{-1}\quad Otherwise.\\
  \end{aligned}
  \right.
\end{equation}
where $\mathcal{Q}$ represents the set of target blocks and $\left| \mathcal{Q} \right|$ represents the size of $\mathcal{Q}$.
Compared to the arithmetic mean, the harmonic mean distinguishes basic blocks that are closer to specific targets, and the more targets a block can reach, the smaller its distance will be, which facilitates covering as many targets as possible.
After a seed $S$ is executed, we analyze its execution trace and compute its code distance. 
We define the seed's code distance as the average of the smallest n block distances along the seed's execution trace, denoted as
\begin{equation}\footnotesize
  \label{eq3}
  CodeDistance(S)=\frac{1}{n}\times\sum_{BB \in \mathcal{N}}{D(BB)}
\end{equation}
where $\mathcal{N}$ represents the set of n basic blocks with the smallest block distances along the execution trace.
The value of n is contingent on the number of basic blocks. 
Notably, we avoid using the average of all block distances as the seed's code distance.
Because the global optimal strategy can lead to discrepancy when there are multiple targets, where seeds reaching the target may exhibit larger distances than those that do not reach any target~\cite{sok}.
We also avoid using the smallest block distance as the seed's code distance, as it fails to effectively differentiate between different seeds.

\textbf{State Distance Evaluation.}
Each code target corresponds to a specific state target, which is a collection of value ranges for state variables.
For a seed $S$, we define its distance to a state target $ST$ as
\begin{equation}\footnotesize
  \label{eq4}
  d_{ST}(S)=\left\{
  \begin{aligned}
  &0 \quad if \ ST == \emptyset,\\
  &\sum_{SV \in \mathcal{M}} R(SV,ST) \quad Otherwise.\\
  \end{aligned}
  \right.
\end{equation}
where $SV$ represents the state variable, $\mathcal{M}$ represents all state variables involved in $ST$.
$R(SV,ST)$ represents the distance of the value $SV$ to the range $ST$ throughout the entire execution trace.
If SV falls within the range $ST$, $R(SV,ST)$ is considered to be 0.
Based on the above definition, for seed $S$, we define its state distance as the harmonic mean of its distances to all state targets, denoted as
\begin{equation}\footnotesize
  \label{eq5}
  StateDistance(S)=\left\{
  \begin{aligned}
  &0 \quad if \ ( \exists ST \in \mathcal{P} \ | \ d_{ST}(S) == 0),\\
  &\left| \mathcal{P} \right|\times[\sum_{ST \in \mathcal{P}}d_{ST}(S)^{-1}]^{-1}\ Otherwise.\\
  \end{aligned}
  \right.
\end{equation}
where $\mathcal{P}$ represents the set of state targets, and  $\left| \mathcal{P} \right|$ represents the size of $\mathcal{P}$.

\textbf{Fitness Metric.}
We evaluate the potential of a seed to trigger vulnerabilities during execution by assigning a score that integrates both its code distance and state distance.
It is defined as
\begin{equation}\footnotesize
  \label{eq6}
  \begin{aligned}
  D=\alpha\times\mathcal{F}(CodeDistance) + (1-\alpha)\times\mathcal{F}(StateDistance)\\
  \end{aligned}
\end{equation}
\begin{equation}\footnotesize
  \label{eq7}
  \begin{aligned}
  SC_{bug}=(D + \beta)^{-1} \\
  \end{aligned}
\end{equation}
where $\alpha$ and $\beta$ are constants less than one, and $\mathcal{F}$ is a normalization function to eliminate the difference in magnitude between the code distance and state distance.
Specifically, we assign $\alpha$ a value of 0.5 and $\beta$ a value of 0.1.
If a seed consists of multiple transactions, the score for the seed is determined by the transaction with the highest score.
The fitness metric also incorporates feedback from branch coverage and data dependencies which are commonly used in graybox fuzzing, resulting in the following representation
\begin{equation}\footnotesize
  \label{eq8}
  \begin{aligned}
  Fitness=\gamma \times SC_{bug} + (1-\gamma)\times(SC_{branch} + SC_{dep})\\
  \end{aligned}
\end{equation}
where $\gamma$ is a constant less than one, $SC_{branch}$ is computed based on the count of newly covered branch edges, and $SC_{dep}$ is computed based on the frequency of state variable writes.

\subsection{Fuzzing Loop}

During the fuzzing loop stage, we use a genetic algorithm~\cite{GA} guided by our proposed fitness metric for seed scheduling. 
We execute contracts using an independent EVM, which we instrument to manage the persistent storage of contract states and provide execution traces for subsequent analysis.

\textbf{Seed Initialization.}
The testing seeds mainly consist of a \textit{function selector}, \textit{calldata}, and \textit{value}. 
The \textit{function selector} determines the function to be invoked and is computed using the contract ABI. 
At the beginning of the test, we initialize two seeds for each function of the contract.
\textit{Calldata} refers to the function parameters. 
We first determine the parameter type according to the ABI. 
For fixed data types, such as \texttt{uint256}, we randomly select values within the valid input range. 
For non-fixed data types, like \texttt{string}, we determine a positive number as the data length and generate an input of that length.
\textit{Value} represents the funds transferred into the contract by the transaction. 
If the function is marked as payable, we attach an appropriate value to the transaction; otherwise, the \textit{value} is 0.

\textbf{Selection.}
We employ the aforementioned fitness metrics to select valuable seeds and allocate more testing resources to them, steering the fuzzing process toward vulnerabilities.
Specifically, it adjusts the seed selection strategy by manipulating the probability of a seed being chosen. 
Seeds with higher fitness are more likely to be selected for subsequent crossover and mutation.
In a fuzzing generation $\mathcal{G}$, we define $P(S)$ as the probability of selecting a seed $S$.
It is computed as
\begin{equation}\footnotesize
  \label{eq9}
  P(S) = Fitness(S)\  / \sum_{G \in \mathcal{G}}Fitness(G).
\end{equation}

\textbf{Crossover.}
We use crossover to generate transaction sequences. 
If two transaction sequences have a Read-After-Write (RAW) dependency, we combine them in the order of RAW. 
Otherwise, the two transaction sequences are combined with a certain probability to generate two longer transaction sequences.

\textbf{Mutation.}
The calldata of the seeds is mutated either randomly or from a mutation pool with a certain probability \textit{P}. 
The mutation pool consists of state targets we identified and values that appeared in previous transactions, and it is continuously expanded during execution. 
This heuristic method increases the likelihood of passing narrow conditional branches.

\textbf{Vulnerability Detection.}
We define a vulnerability as being discovered by analyzing the execution trace to see if it violates the test oracles. 
Our test oracles are created following those used in previous works such as \textsc{ContractFuzzer}~\cite{contractfuzzer} and \textsc{Smartian}~\cite{Smartian}.

\section{Implementation.}
\label{Implementation}
\textsc{Vulseye} is implemented in Python, and we use py-evm~\cite{py-evm}, the official Python implementation of EVM, for the execution of smart contracts.
We implement the static analysis phase of \textsc{Vulseye} in an over-approximate way, so that more targets can be tested during the fuzzing phase, reducing the false negatives.
Due to the uncertainty of the interactive objects when contracts initiate external calls, ensuring the success of every external call during testing is challenging~\cite{xFuzz}. 
To address this, we instrument py-evm to ensure that external calls within contracts consistently return appropriate values, thus preventing execution failures and facilitating cross-contract testing.
We modularize our test oracle to facilitate easy extension for supporting more vulnerability detections.

\begin{figure*}[t!]
  \centering

  \subfigure[Distance of \textsc{Vulseye}]{
    \begin{minipage}[b]{0.22\textwidth}
      \includegraphics[width=1\textwidth]{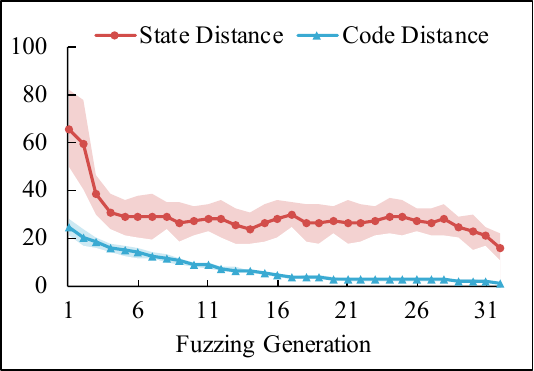} 
    \end{minipage}
  }
      \subfigure[Distance of configuration A]{
        \begin{minipage}[b]{0.22\textwidth}
          \includegraphics[width=1\textwidth]{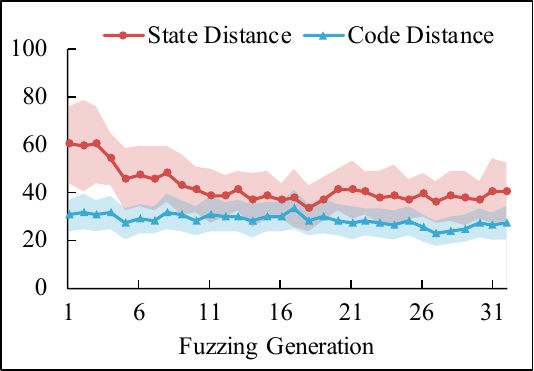}
        \end{minipage}
      }
  \subfigure[Distance of configuration B]{
    \begin{minipage}[b]{0.22\textwidth}
      \includegraphics[width=1\textwidth]{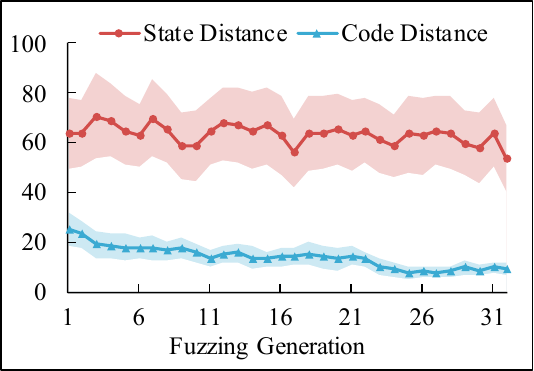} 
    \end{minipage}
  }
      \subfigure[Distance of \textsc{ConFuzzius}]{
        \begin{minipage}[b]{0.22\textwidth}
       \includegraphics[width=1\textwidth]{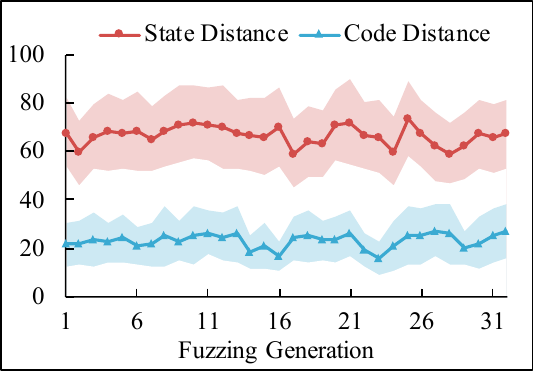}
        \end{minipage}
      }
  \caption{Tendency of code/state distance on different ablation configurations.}
  \label{figure 5}
  \vspace{-5pt}

\end{figure*}

\begin{figure*}[t!]
  \centering

  \subfigure[Code coverage of $B_2$]{
    \begin{minipage}[b]{0.22\textwidth}
      \includegraphics[width=1\textwidth]{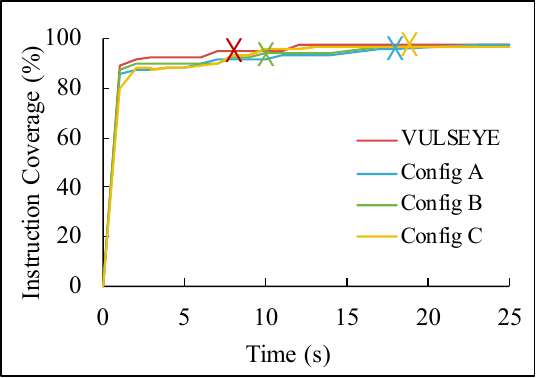} 
    \end{minipage}
  }
      \subfigure[Code coverage of $B_3$]{
        \begin{minipage}[b]{0.22\textwidth}
          \includegraphics[width=1\textwidth]{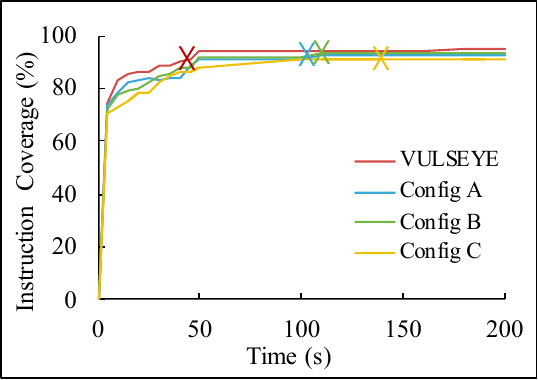}
        \end{minipage}
      }
  \subfigure[Code coverage of $B_4$]{
    \begin{minipage}[b]{0.22\textwidth}
      \includegraphics[width=1\textwidth]{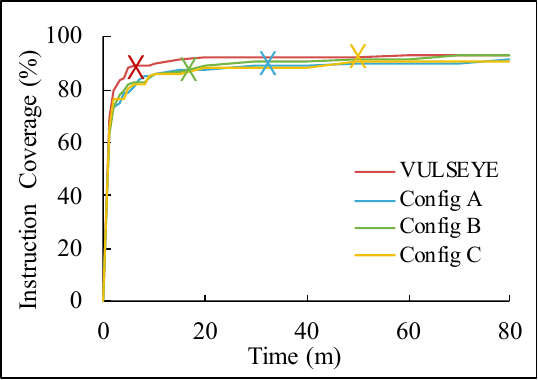} 
    \end{minipage}
  }
      \subfigure[Code coverage of $B_5$]{
        \begin{minipage}[b]{0.22\textwidth}
       \includegraphics[width=1\textwidth]{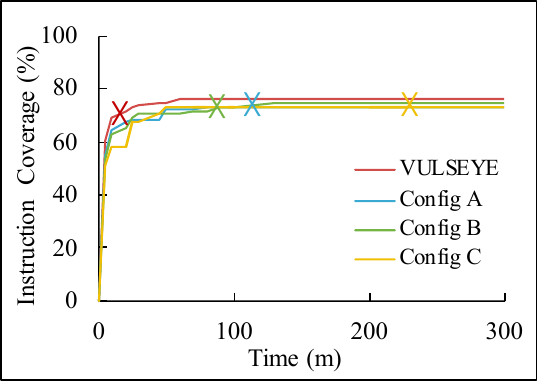}
        \end{minipage}
      }
  \caption{Tendency of code coverage on different benchmarks. The \texttt{X} indicates when and at what coverage the test target is reached.}
  \label{figure 6}

  \vspace{-5pt}

\end{figure*}

\section{Experiments}
\label{Experiments}

\textbf{Rearch Questions.}
\label{Rearch Questions.}
We conduct experiments to answer the following four questions.
Our testing environment is comprised of a server with a 16-core Intel(R)-Xeon(R)-Gold-5218 CPU \( @\)2.30 GHz, 340GB of RAM, and the Ubuntu 18.04 LTS operating system.

\begin{itemize}[leftmargin=4mm, itemindent=0mm]

  \item \textbf{RQ1:} How effective is \textsc{Vulseye} in guiding testing toward specific targets, and how significantly do the code and state targets contribute to its performance?
    
  \item \textbf{RQ2:} How effective is \textsc{Vulseye} in vulnerability detection compared to state-of-the-art approaches.
  
  \item \textbf{RQ3:} How does \textsc{Vulseye} perform on code coverage?
  
  \item \textbf{RQ4:} How does \textsc{Vulseye} perform on testing real-world smart contracts?
  
  \end{itemize}

\subsection{RQ1: Testing Specific Targets}
\label{RQ1}
We validate the efficacy of \textsc{Vulseye} in testing specific code areas and create three ablations to evaluate the impact of both code and state targets that we introduced.

\textbf{Dataset.}
We craft five benchmarks for RQ1 as existing datasets don't meet our needs and cannot contribute to a fair evaluation.
We require: i) each contract to contain a vulnerability in a suitable branch to serve as a test target; ii) sequentially increasing difficulty in reaching these targets; iii) sufficiently challenging branch conditions to extend fuzzing time for observation.
To achieve this, each contract is instrumented with hard-to-trigger branches by intentionally imposing restrictions on contract states. 
From these branches within each contract, we select one to inject a suicidal vulnerability as the testing target. 
The difficulty of reaching these targets escalates sequentially across these five benchmarks, from $B_1$ to $B_5$.
Each benchmark consists of 10 contracts, and their average results are used to mitigate variability.

\begin{table}[t!]
  \caption{Comparison of time to targets\tnote{*}.}
  \label{table 3}
  \scalebox{0.94}{
  \begin{threeparttable}

  \begin{tabular}{c|r|r|r|r|r|r|r}
  \multicolumn{1}{c|}{\textbf{}} & \multicolumn{1}{c|}{\textbf{T$_V$}} & \multicolumn{1}{c|}{\textbf{T$_A$}} & \multicolumn{1}{c|}{\textbf{T$_B$}} & \multicolumn{1}{c|}{\textbf{T$_C$}} & \multicolumn{1}{c|}{$\frac{\textbf{T$_A$}}{\textbf{T$_V$}}$} & \multicolumn{1}{c|}{$\frac{\textbf{T$_{B}$}}{\textbf{T$_V$}}$} & \multicolumn{1}{c}{$\frac{\textbf{T$_{C}$}}{\textbf{T$_V$}}$} \\ \hline
  \textbf{B$_1$}                    & \textbf{2.30}                                & 4.35                               & 2.71                               & 4.14                            & 1.89                                  & 1.18                                   & 1.80                                 \\
  \textbf{B$_2$}                    & \textbf{7.70}                                & 17.79                              & 9.32                                 & 18.25                             & 2.31                                  & 1.21                                  & 2.37                                \\
  \textbf{B$_3$}                    & \textbf{42.90}                                & 98.67                              & 110.25                             & 138.57                               & 2.30                                 & 2.57                                 & 3.23                                \\
  \textbf{B$_4$}                    & \textbf{494.42}                                & 1888.68                              & 1112.45                               & 2882.47                               & 3.82                                 & 2.25                                  & 5.83                                \\
  \textbf{B$_5$}                   & \textbf{1618.34}                                & 6667.56                               & 5502.36                               & 13691.16                               & 4.12                                 & 3.40                                  & 8.46                                 \\ \hline
  \end{tabular}
  \begin{tablenotes}
    \footnotesize
    \item[*]B$_1$-B$_5$ are five benchmarks. T$_V$ is the time (seconds) \textsc{Vulseye} takes to reach targets, and T$_{A}$, T$_{B}$, T$_{C}$ are the time of configuration A, B, and C.
  \end{tablenotes}
  \end{threeparttable}}
  \end{table}

\textbf{Ablation Study.}
We test \textsc{Vulseye} on these five benchmarks, calculating the average time consumed to reach these targets.
We create three ablations, excluding the guidance of either code targets, state targets, or both (designated as configurations A, B, and C respectively).
We conducted this experiment five times, and the results are presented in Table~\ref{table 3}.
The first column denotes the five benchmarks, while the second to fifth columns display the average time taken by \textsc{Vulseye}, configurations A, B, and C, respectively, to reach the targets.
Columns six to eight present the speed-up factor achieved by \textsc{Vulseye}.
As depicted in Table~\ref{table 3}, \textsc{Vulseye} achieves the shortest time to reach the targets in all five benchmarks compared to the ablations.
Notably, the speed-up factor is more pronounced in the tests involving more intricate benchmarks, showcasing a more substantial efficiency improvement, reaching up to 8.46$\times$ when compared with configuration C.
Moreover, configurations A and B take more time than \textsc{Vulseye} (1.18 to 4.12$\times$), suggesting that both code target and state target contribute to \textsc{Vulseye}'s ability in testing specific code areas.

To visually illustrate how \textsc{Vulseye} guides testing toward these targets, we depict the tendency of code and state distance during the test of a contract within $B_4$ in Figure~\ref{figure 5}, with values averaged over all seeds within a fuzzing generation.
In Figure~\ref{figure 5}a, we note a consistent decrease in the state and code distances of \textsc{Vulseye} with the increase in fuzzing generations, exhibiting minimal standard deviations.
In contrast, configurations A and B exhibit larger fluctuations and standard deviations in code distance and state distance, respectively. 
Configuration C, lacking guidance in both code and state space, exhibits random changes in both state and code distance in Figure~\ref{figure 5}d.
Additionally, we depict the tendency of average code coverage across different benchmarks in Figure~\ref{figure 6}. 
It can be seen that \textsc{Vulseye} outperforms the ablations in terms of coverage while reaching the targets faster.

\begin{center}
  \fcolorbox{black}{gray!10}{\parbox{.97\linewidth}{\vspace{0.9ex} \leftskip=0.6em \rightskip=0.6em  \textbf{Answer to RQ1:} With the guidance of code and state targets we introduced, \textsc{Vulseye} directs testing toward specific code areas effectively. It reaches challenging targets up to 8.4$\times$ faster than the baselines.\vspace{0.9ex}}}
\end{center}

\subsection{\scalebox{0.95}{RQ2: Effectiveness of Vulnerability Detection}}
\label{RQ2}

To answer RQ2, we conduct comparative experiments between \textsc{Vulseye} and SOTA approaches on a ground-truth dataset consisting of 42 previously exploited projects.

\textbf{Dataset.}
Our second dataset is derived from previously exploited projects curated by Durieux et al~\cite{durieux2020empirical} and Torres et al~\cite{ConFuzzius}.
However, these contracts lack the desired quantity and complexity for certain vulnerability types. 
To address this limitation, we have expanded and enriched these contracts to augment both their quantity and complexity.
As a result, our dataset encompasses vulnerabilities of various types, including ether leaking (EL), block dependency (BD), reentrancy (RE), controlled delegatecall (CD), dangerous delegatecall (DD), suicidal (SD), and locking ether (LE).

\textbf{Baselines Selection.}
In the experiment of RQ2, we compare \textsc{Vulseye} to SOTA graybox fuzzers that are publicly available, including \textsc{ConFuzzius}~\cite{ConFuzzius}, \textsc{sFuzz}~\cite{sFuzz}, \textsc{Smartian}~\cite{Smartian}, \textsc{IR-Fuzz}~\cite{Rethinking}, and \textsc{ITYFuzz}~\cite{ITYFuzz}.
We choose \textsc{ConFuzzius} because it is the first hybrid fuzzer designed for smart contracts.
We choose \textsc{sFuzz} because its fuzzing strategy is inspired by AFL~\cite{AFL}, a highly effective and widely-used fuzzer for C programs.
We choose \textsc{ITYFuzz} because, to our knowledge, it is the first tool to employ snapshot technology in smart contract fuzzing.
We choose \textsc{Smartian} and \textsc{IR-Fuzz} since they represent the latest advancements in research and assert superior performance compared to prior works.

\textbf{Vulnerability Detection.}
We applied \textsc{Vulseye} and SOTA tools on this ground-truth dataset.
We performed five runs for each contract, employing a fuzzing timeout of 15 minutes for contracts with a code size below 200 lines and extending it to 30 minutes for larger contracts.
The experimental results are presented in Table~\ref{table 4}, where the first column indicates the contract, the second column specifies the vulnerability type within the contract, and the third column identifies the SWC ID according to the Smart Contract Weakness Classification~\cite{swc}. 
The average time to detect vulnerabilities is recorded in columns four through nine.

As shown in Table~\ref{table 4}, \textsc{Vulseye} successfully detected 42 out of 42 vulnerabilities, \textsc{ConFuzzius} detected 38 vulnerabilities, and \textsc{Smartian} detected 29 vulnerabilities.
\textsc{sFuzz} and \textsc{IR-Fuzz} lacked test oracles for suicidal and ether leak, and exhibited poor performance in detecting reentrancy.
As a result, they only detected 16 and 14 out of 42 vulnerabilities, respectively.
\textsc{ITYFuzz} lacked test oracles for delegatecall types and lock Ether, identifying 17 vulnerabilities.
In terms of the time consumed to discover vulnerabilities, \textsc{Vulseye} outperforms the other tools significantly.
Particularly, our tool achieves the shortest time for the detection of 34 out of the 42 vulnerabilities (account for \textbf{81\%}). 
Notably, this efficiency improvement is more significant especially for contracts larger than 200 lines (13 out of 14 vulnerabilities, account for \textbf{93\%}).
This is due to \textsc{Vulseye}'s prioritization of testing resources in vulnerable areas.

\begin{table}[t!]
  \caption{Comparison of time to vulnerabilities\tnote{*}.}
  \label{table 4}
  \scalebox{0.8}{
  \begin{threeparttable}

  \begin{tabular}{@{}c|c|c|r|r|r|r|r|r@{}}
  \textbf{CID} & \textbf{BID} & \textbf{SWC-ID}  & \multicolumn{1}{c|}{\textbf{VE}} & \multicolumn{1}{c|}{\textbf{CF}} & \multicolumn{1}{c|}{\textbf{SF}} & \multicolumn{1}{c|}{\textbf{IF}} & \multicolumn{1}{c|}{\textbf{IT}} & \multicolumn{1}{c}{\textbf{ST}} \\ \hline
  1            & EL           & SWC-105                      & 16.87                              & 3.56                              & -                              & -                              & \textbf{1.02}                          & 7.70                \\
  2            & EL           & SWC-105                      & \textbf{6.63}                              & 26.10                              & -                              & -                              & 10.30               & 900.00                           \\
  3            & EL           & SWC-105                     & 13.87                              & \textbf{2.24}                              & -                              & -                              &   2.25                & 4.96                       \\
  4            & EL           & SWC-105                     & \textbf{2.86}                              & 15.39                              & -                              & -                              &  3.63                        & 900.00                       \\
  5            & EL           & SWC-105                     & \textbf{0.15}                              & 4.02                              & -                              & -                              &   1.11                & 30.12                     \\
  6            & EL           & SWC-105                     & \textbf{23.8}                              & 185.60                              & -                              & -                              &  32.95             & 1800.00                         \\
  7            & BD           & SWC-120                      & \textbf{0.64}                              & 4.21                              & 10.38                              & 55.50                              &  0.90         & 6.57                          \\
  8            & BD           & SWC-120                      & \textbf{1.01}                              & 2.33                              & 900.00                              & 900.00                              &  900.00       & 4.79                           \\
  9            & BD           & SWC-120                     & \textbf{1.99}                              & 64.50                              & 5.50                              & 5.63                              &  900.00          & 18.94                  \\
  10           & BD           & SWC-120                     & \textbf{0.11}                              & 3.23                              & 15.37                              & 77.68                              &  900.00         & 13.61                   \\
  11           & BD           & SWC-120                     & 9.15                              & 32.11                               & 5.89                              & 369.22                              &  \textbf{4.56}        & 68.42                    \\
  12           & BD           & SWC-120                     & \textbf{1.06}                              & 37.24                              & 355.14                              & 488.45                              &  1800.00     & 539.67                       \\
  13           & RE           & SWC-107                      & \textbf{2.05}                              & 2.14                              & 21.33                              & 9.12                              &   4.55       & 200.50                    \\
  14           & RE           & SWC-107                      & \textbf{6.27}                              & 15.02                              & 900.00                              & 900.00                              &  9.03     & 87.43                       \\
  15           & RE           & SWC-107                    & \textbf{16.89}                              & 26.33                              & 900.00                              & 900.00                              &  900.00     & 109.11                       \\
  16           & RE           & SWC-107                    & \textbf{3.06}                              & 8.12                             & 900.00                              & 900.00                              &  19.48       & 900.00                     \\
  17           & RE           & SWC-107                     & \textbf{8.73}                              & 23.45                              & 1800.00                              & 1800.00                              &  1800.00    & 55.34                        \\
  18           & RE           & SWC-107                     & \textbf{211.29}                              & 377.72                              & 1800.00                              & 1800.00                              & 1800.00 & 1800.00                            \\
  19           & CD           & SWC-112                      & 5.45                              & 3.30                              & 5.29                              & 24.06                              & -            & \textbf{2.23}                \\
  20           & CD           & SWC-112                      & 4.12                              & \textbf{2.50}                              & 5.60                              & 19.23                              &  -           & 5.28                \\
  21           & CD           & SWC-112                    & \textbf{42.21}                              & 224.34                              & 900.00                              & 476.45                              &  -      & 48.33                      \\
  22           & CD           & SWC-112                    & 99.09                              & 7.49                              & \textbf{5.12}                              & 64.22                              &  -            & 6.69                \\
  23           & CD           & SWC-112                     & \textbf{3.06}                              & 69.68                              & 10.88                              & 343.78                              &  -        & 11.61                   \\
  24           & CD           & SWC-112                     & \textbf{5.45}                              & 29.97                              & 35.43                              & 1800.00                              & -         & 26.78                     \\
  25           & DD           & SWC-112                      & \textbf{3.92}                              & 6.89                              & 5.67                              & 900.00                              & -         & 900.00                    \\
  26           & DD           & SWC-112                      & \textbf{0.15}                              & 2.02                              & 5.30                              & 13.10                              & -            & 2.29                 \\
  27           & DD           & SWC-112                    & \textbf{126.59}                              & 900.00                              & 900.00                              & 900.00                              &  -    & 900.00                        \\
  28           & DD           & SWC-112                    & 24.53                              & 34.23                              & 5.98                              & \textbf{4.54}                              &  -            & 5.32               \\
  29           & DD           & SWC-112                     & \textbf{1.25}                              & 53.30                              & 5.70                              & 3.20                              &  -            & 2.71                \\
  30           & DD           & SWC-112                     & \textbf{12.13}                              & 31.54                              & 30.31                              & 1800.00                              &  -      & 25.42                      \\
  31           & SD           & SWC-106                      & \textbf{1.01}                              & 2.21                              & -                              & -                              &  1.33                  & 1.88          \\
  32           & SD           & SWC-106                      & 1.42                              & 2.09                              & -                              & -                              &  \textbf{0.89}                           & 1.34          \\
  33           & SD           & SWC-106                    & \textbf{0.49}                              & 4.18                              & -                              & -                              &  1.33                    & 6.61        \\
  34           & SD           & SWC-106                    & \textbf{4.47}                              & 14.71                              & -                              & -                              &  5.78                   & 5.92         \\
  35           & SD           & SWC-106                     & \textbf{10.54}                              & 19.66                              & -                              & -                              &  31.26               & 226.45             \\
  36           & SD           & SWC-106                     & \textbf{107.29}                              & 382.64                              & -                              & -                              &  198.14             & 159.03               \\
  37           & LE           & SWC-132                      & \multicolumn{1}{r|}{\Checkmark}                             & \multicolumn{1}{r|}{\XSolidBrush}                              & \multicolumn{1}{r|}{\XSolidBrush}                              & \multicolumn{1}{r|}{\XSolidBrush}                              & \multicolumn{1}{r|}{\XSolidBrush}    & \multicolumn{1}{r}{\XSolidBrush}                         \\
  38           & LE           & SWC-132                      & \multicolumn{1}{r|}{\Checkmark}                              & \multicolumn{1}{r|}{\Checkmark}                              & \multicolumn{1}{r|}{\XSolidBrush}                              & \multicolumn{1}{r|}{\XSolidBrush}                              & \multicolumn{1}{r|}{\XSolidBrush}     & \multicolumn{1}{r}{\XSolidBrush}                       \\
  39           & LE           & SWC-132                    & \multicolumn{1}{r|}{\Checkmark}                              & \multicolumn{1}{r|}{\XSolidBrush}                              & \multicolumn{1}{r|}{\XSolidBrush}                              & \multicolumn{1}{r|}{\XSolidBrush}                              & \multicolumn{1}{r|}{\XSolidBrush}     & \multicolumn{1}{r}{\XSolidBrush}                        \\
  40           & LE           & SWC-132                    & \multicolumn{1}{r|}{\Checkmark}                              & \multicolumn{1}{r|}{\Checkmark}                              & \multicolumn{1}{r|}{\XSolidBrush}                              & \multicolumn{1}{r|}{\XSolidBrush}                              & \multicolumn{1}{r|}{\XSolidBrush}       & \multicolumn{1}{r}{\XSolidBrush}                      \\
  41           & LE           & SWC-132                     & \multicolumn{1}{r|}{\Checkmark}                              & \multicolumn{1}{r|}{\Checkmark}                              & \multicolumn{1}{r|}{\XSolidBrush}                              & \multicolumn{1}{r|}{\XSolidBrush}                              & \multicolumn{1}{r|}{\XSolidBrush}      & \multicolumn{1}{r}{\XSolidBrush}                       \\
  42           & LE           & SWC-132                     & \multicolumn{1}{r|}{\Checkmark}                              & \multicolumn{1}{r|}{\XSolidBrush}                              & \multicolumn{1}{r|}{\XSolidBrush}                              & \multicolumn{1}{r|}{\XSolidBrush}                              & \multicolumn{1}{r|}{\XSolidBrush}    & \multicolumn{1}{r}{\XSolidBrush}                         \\ \bottomrule
  \end{tabular}
  \begin{tablenotes}
    \footnotesize
    \item[*]VE represents \textsc{Vulseye}, CF represents \textsc{ConFuzzius}, SF represents \textsc{sFuzz}, IF represents \textsc{IR-Fuzz}, IT represents \textsc{ITYFuzz}, and ST represents \textsc{Smartian}. Since locking ether (LE) vulnerabilities are only reported upon the fuzzing process is terminated, \Checkmark represents successful detections and \XSolidBrush represents failed detections within the specified time limit. 
  \end{tablenotes}
  \end{threeparttable}}

  \end{table}

\begin{center}
\fcolorbox{black}{gray!10}{\parbox{.97\linewidth}{\vspace{0.9ex} \leftskip=0.6em \rightskip=0.6em  \textbf{Answer to RQ2: } \textsc{Vulseye} outperforms SOTA tools in smart contract vulnerability detection. It exploited all vulnerabilities in the 42 projects and exhibited the fastest detection time for 81\% of them; this figure rises to 93\% when the project is larger.\vspace{0.9ex}}}
\end{center}

\subsection{RQ3: Code Coverage}
\label{RQ3}

\begin{figure}[t!]
	\centering
	\includegraphics[width=3.3in]{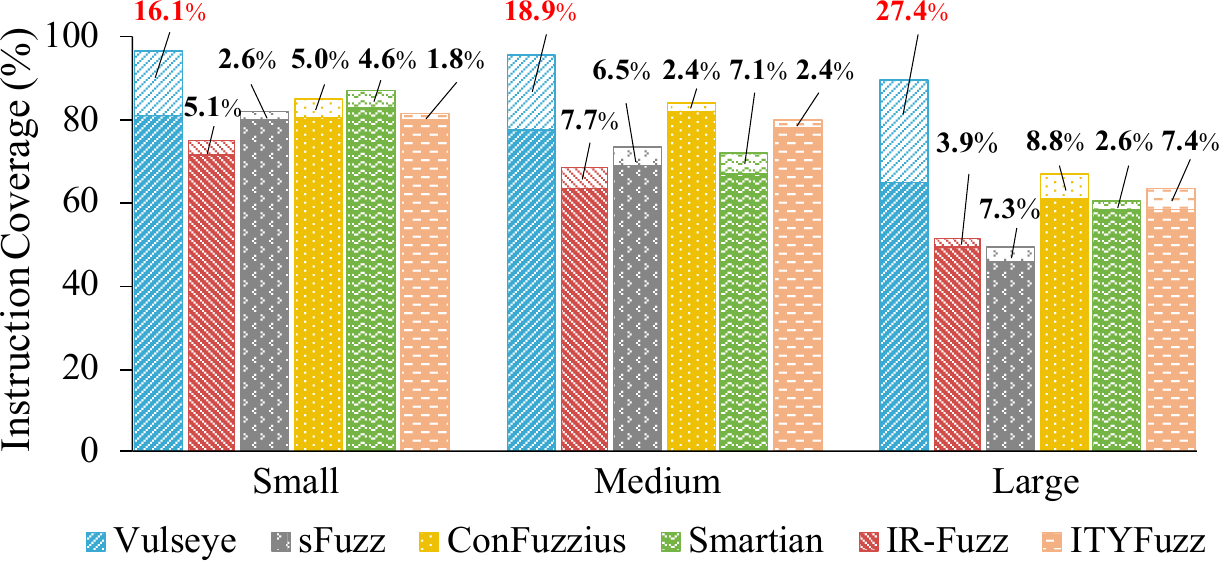}
	\caption{Instruction coverage upon vulnerability discovery (dark shading) and upon fuzzing is terminated (dark shading and light shading).}
	\label{figure 7}
  \end{figure}

We categorized the contracts in RQ2 according to their size into three groups: small, medium, and large, and recorded the average instruction coverage (a form of code coverage) achieved by the six tools.
As shown in Figure~\ref{figure 7}, \textsc{Vulseye} achieved the final instruction coverage of 96.9\%, 95.5\%, and 89.9\%, surpassing all other tools.
While achieving notable results in final instruction coverage, \textsc{Vulseye} does not solely prioritize maximizing this metric.
The light shading in Figure~\ref{figure 7} represents the difference between the final instruction coverage and the coverage upon vulnerability discovery.
A larger light shading area indicates a lower instruction coverage requirement for vulnerability detection.
Comparing the light shading areas, they exhibit the largest proportion in \textsc{Vulseye} (16.1\%, 18.9\%, 27.4\%) within all three groups.
This suggests that with the fuzzing strategy we employed, \textsc{Vulseye} doesn't require achieving such a high code coverage to expose vulnerabilities, leading to a significant improvement in vulnerability discovery efficiency.
It is worth noting that, although the code coverage at vulnerability discovery (dark shading) of comparison tools is lower than that of \textsc{Vulseye} in Figure~\ref{figure 7}, it's due to their limitation in the notably lower final code coverage.
However, their proportion of the dark shading areas remains higher than that of \textsc{Vulseye} (the lower the better).

\begin{center}
  \fcolorbox{black}{gray!10}{\parbox{.97\linewidth}{\vspace{0.9ex} \leftskip=0.6em \rightskip=0.6em  \textbf{Answer to RQ3:} \textsc{Vulseye} achieves the highest code coverage among the SOTA tools. It doesn't necessitate as high code coverage as SOTA tools to detect vulnerabilities, highlighting its significant advantage of prioritizing testing resources on vulnerable code and states.\vspace{0.9ex}}}
\end{center}

\subsection{RQ4: Performance on Real-World Contracts}
\label{RQ4}

To answer RQ4, we employed \textsc{Vulseye} on 42,738 real-world smart contracts for bug detection and compared it with other tools.
Additionally, we collected top 50 DApps from Ethereum for testing and found previously unknown bugs.

\textbf{Finding Bugs in 42,738 Real-World Smart Contracts.}
We obtained the dataset from the study of Durieux et al~\cite{durieux2020empirical}, consisting of 47,587 real-world smart contracts extracted from Ethereum.
After excluding contracts that couldn't be compiled by our compiler, this dataset comprised 42,738 real-world contracts.
The source code for all contracts in this dataset is available on Etherscan~\cite{Etherscan}.

\begin{figure}[t]
	\centering
	\includegraphics[width=2.8in]{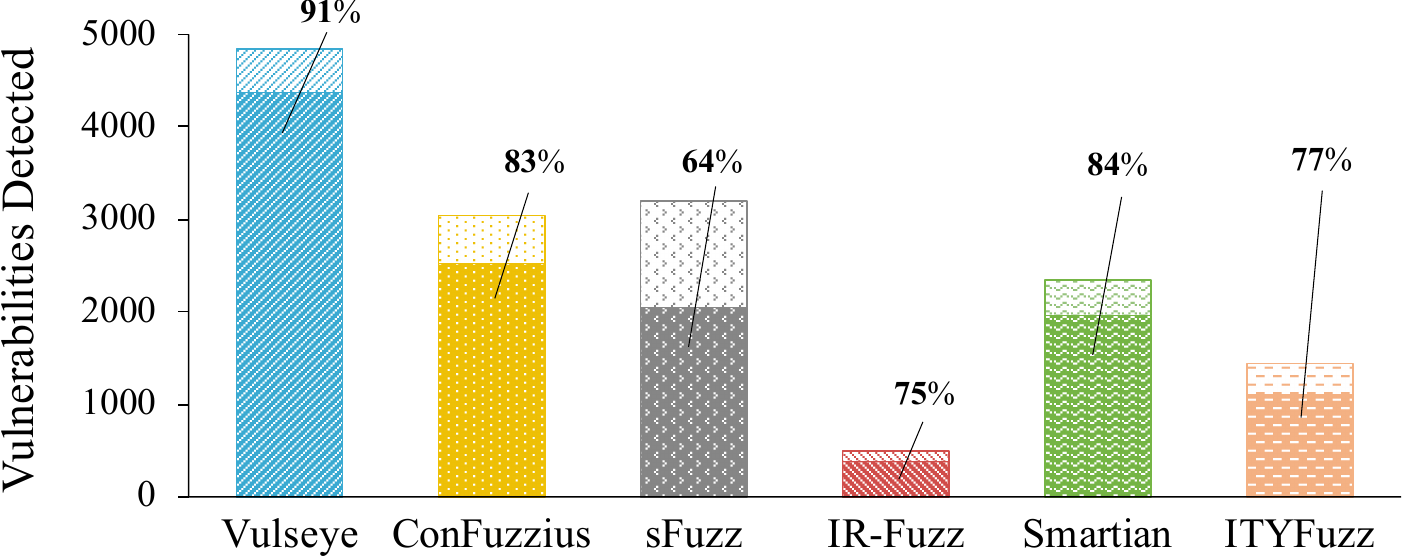}
	\caption{The total number of identified vulnerabilities reported by each method (dark shading and light shading) and the true positive rates obtained from manual inspection (dark shading).}
	\label{figure 8}
  \end{figure}

We applied \textsc{Vulseye} and other comparison tools on these contracts.
We set the maximum number of test cases for each smart contract as 2,000, which we consider sufficient for most exploits.
To further evaluate the performance of \textsc{Vulseye}, we examine the identified vulnerable contracts manually to see whether they are true positives or not.
Due to the large number of reported vulnerabilities, we are unable to check all of them.
Instead, for each tool, we randomly sample 350 vulnerable contracts for manual verification (50 contracts for each vulnerability type).
In cases where the reported results involve less than 50 contracts, all of them will be subject to manual verification.
Table~\ref{table 5} reveals that \textsc{Vulseye} identified 4,845 vulnerabilities, surpassing other tools by up to 9.7$\times$.
Additionally, the overall true positive rate for \textsc{Vulseye} in manual checking is the highest.
Specifically, \textsc{Vulseye} detected more vulnerabilities in 4 out of 7 vulnerability categories and demonstrated the highest true positive rate among the 6 vulnerability categories.
Figure~\ref{figure 8} provides an overview of the experimental results, clearly demonstrating that \textsc{Vulseye} outperforms other tools in both vulnerability discovery quantity and true positive rate.

\begin{table}[t!]
	\caption{Comparison of detected vulnerabilities and true positive rate.}
	\label{table 5}
	\setlength{\tabcolsep}{3.29pt}
	\scalebox{0.8}{
	\begin{tabular}{c|rr|rr|rr|rr|rr|rr}
	\multirow{2}{*}{\textbf{BID}} & \multicolumn{2}{c|}{\textbf{\textsc{Vulseye}}}        & \multicolumn{2}{c|}{\textbf{\textsc{ConFuzzius}}}         & \multicolumn{2}{c|}{\textbf{\textsc{sFuzz}}}              & \multicolumn{2}{c|}{\textbf{\textsc{IR-Fuzz}}}            & \multicolumn{2}{c|}{\textbf{\textsc{Smartian}}}   & \multicolumn{2}{c}{\textbf{\textsc{ITYFuzz}}}         \\
								  & \multicolumn{1}{c}{\#} & \multicolumn{1}{c|}{tp} & \multicolumn{1}{c}{\#} & \multicolumn{1}{c|}{tp} & \multicolumn{1}{c}{\#} & \multicolumn{1}{c|}{tp} & \multicolumn{1}{c}{\#} & \multicolumn{1}{c|}{tp} & \multicolumn{1}{c}{\#} & \multicolumn{1}{c|}{tp} & \multicolumn{1}{c}{\#} & \multicolumn{1}{c}{tp}\\ \hline
	EL                            & 408                    & 88\%                    & 351                    & 86\%                    & 0                      & N/A                     & 0                      & N/A                     & 387                    & 86\%           & 838        & 78\%\\
	BD                            & 3720                   & 100\%                   & 2032                   & 78\%                    & 2737                   & 100\%                   & 343                    & 84\%                    & 1825                   & 80\%           & 455        & 84\%\\
	RE                            & 426                    & 92\%                    & 397                    & 82\%                    & 278                    & 78\%                    & 99                     & 86\%                    & 25                     & 68\%           & 132       & 64\%\\
	CD                            & 12                     & 100\%                   & 5                      & 100\%                   & 36                     & 33\%                    & 10                     & 70\%                    & 1                      & 0\%            & 0      & N/A\\
	DD                            & 26                     & 100\%                   & 7                      & 100\%                   & 66                     & 42\%                    & 23                     & 48\%                    & 8                      & 100\%          & 0       & N/A\\
	SD                            & 87                     & 100\%                   & 131                    & 80\%                   & 0                      & N/A                     & 0                      & N/A                     & 94                     & 92\%            & 22       & 86\%\\
	LE                            & 166                    & 64\%                    & 113                    & 84\%                    & 92                     & 58\%                    & 25                     & 60\%                    & 0                      & N/A            & 0        & N/A\\ \hline
	\textbf{Total}                & \textbf{4,845}                   & \textbf{91\%}                    & 3036                   & 83\%                    & 3209                   & 64\%                    & 500                    & 75\%                    & 2340                   & 84\%      & 1447                   & 77\%              \\ \hline
	\end{tabular}}
	\end{table}

\textbf{Case Study.}
To provide further insight on how the fuzzing strategies implemented in \textsc{Vulseye} enhance bug detection and surpass existing fuzzers, we present a case study of a contract named \textit{THE\_BANK}\footnote{0xCB6fe98097fE7D6E00415Bb6623D5fC3EFFA4E83}, which contains a reentrancy vulnerability. 
This vulnerability was detected by \textsc{Vulseye} within the aforementioned real-world dataset but remained undiscovered by other tools.
As shown in Figure~\ref{figure Case Study}, to trigger the reentrancy in Line 2, specific values of the state variables \texttt{acc.balance} and \texttt{acc.unlockTime} are required, which are influenced by other functions.
During testing of \textit{THE\_BANK} by \textsc{Vulseye} and other fuzzers, we recorded the average code and state distances to the reentrancy vulnerability in each fuzzing generation, as depicted in Figure~\ref{figure Case Study2}.
\textsc{Vulseye} showed a consistent decrease in both code and state distances, uncovering the vulnerability by the 37th test generation. 
It demonstrated that \textsc{Vulseye} successfully identified this vulnerable location, determined the necessary state conditions, and set corresponding targets to guide the fuzzing. 
In contrast, other fuzzers exhibited erratic changes in distance and failed to detect the vulnerability in the limited time.
Their inefficient allocation of testing resources hinders their ability to set the contract state appropriately and exploit the vulnerability.

\textbf{False Positives.}
During manual verification, we summarized several reasons for false positives reported by \textsc{Vulseye}.
For ether leaking (EL), the false positives attribute to contracts designed for lotteries, where the outward sending of ether is the original intention of the contract designer and should not be considered a vulnerability.
For reentrancy (RE), the false positives result from instances where contracts are immune to reentrancy attacks due to inherent reentrancy protection or restrictions on external calls. 
In fact, it is non-trivial to automatically conduct reentrancy attacks on smart contracts.
False positives in locking ether (LE) stem from contracts that cannot directly send funds but instead employ delegatecall to engage other contracts in fund transfers.

\textbf{Finding 0-Day Bugs in Top DApps.}
We collected the top 50 popular DApps listed on BitDegree~\cite{bitdegree}, a famous Web3 platform.
These DApps primarily include projects in DeFi, gaming, and gambling, involving millions of transactions on the blockchain, and we utilized \textsc{Vulseye} to test them.
After five runs for each DApp, \textsc{Vulseye} reported a total of 16 warnings involving 6 projects.
After manual verification, we identified 11 previously unknown bugs related to reentrancy, controlled delegatecall, and block dependencies within 4 projects.
The estimated balance of these four projects on BitDegree is about 2,500,000 USD, and these vulnerabilities put these funds at risk.
We have responsibly reported these vulnerabilities to the corresponding authorities.

\begin{center}
  \fcolorbox{black}{gray!10}{\parbox{.97\linewidth}{\vspace{0.9ex} \leftskip=0.6em \rightskip=0.6em  \textbf{Answer to RQ4:} \textsc{Vulseye} is effective at identifying vulnerabilities in real-world scenarios, outperforming existing tools by reporting more vulnerabilities and achieving the highest true positive rate.\vspace{0.9ex}}}
\end{center}

\begin{figure}[t!]
	\centering
	\begin{minipage}{0.45\textwidth}
	\begin{lstlisting}[language = Solidity]
if(acc.balance>=MinSum && acc.balance>=_am && now>acc.unlockTime){
	if(msg.sender.call.value(_am)()){acc.balance-=_am;} }
	\end{lstlisting}
	\end{minipage}
  \vspace{-5pt}
	\caption{Code Snippet of \textit{THE\_BANK}}
	\label{figure Case Study}
  \vspace{-10pt}
	\end{figure}

	\begin{figure}[t!]
		\centering
	  
	\subfigure[Code Distance]{
		\begin{minipage}[b]{0.22\textwidth}
		\includegraphics[width=1\textwidth]{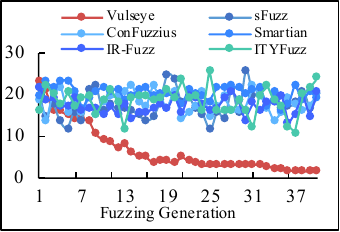} 
		\end{minipage}
	}
	\subfigure[State Distance]{
	\begin{minipage}[b]{0.22\textwidth}
	\includegraphics[width=1\textwidth]{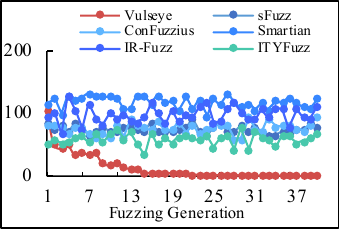}
	\end{minipage}
	}
  \vspace{-5pt}
	\caption{Tendency of code/state distance during testing of \textit{THE\_BANK}.}
	\label{figure Case Study2}
	\vspace{-8pt}
	\end{figure}

\subsection{Threats to Validity}

First, the performance of \textsc{Vulseye} varies with the choice of the initial seeds.
To ensure fairness, we used the same initial seeds in our comparative experiments with SOTA works, but other initial seeds or benchmarks may produce different results.
We have open-sourced \textsc{Vulseye} to facilitate further evaluation in other studies.
Second, the false positive rate in RQ4 may not be entirely accurate. 
Due to the large number of contracts, we randomly selected 350 contracts from the positive results and manually checked them to estimate the tool's false positive rate for the entire dataset.
Another threat to validity relates to the type of vulnerabilities reported by \textsc{ITYFuzz} in the comparative experiment. 
Since \textsc{ITYFuzz}'s outputs do not directly indicate the vulnerability type, we determine it by manual inspection.

\section{Related Work}
\label{Related Work}
We briefly review related works on smart contract vulnerability discovery and techniques of directed-graybox fuzzing.
\subsection{Smart Contracts Vulnerability Discovery}

\textbf{Smart Contracts Fuzzing.}
\textsc{Vulseye} is closely related to work on smart contracts fuzzing~\cite{contractfuzzer,Harvey,sFuzz,ConFuzzius,onchainFuzz,Effectively}.
\textsc{ContractFuzzer}~\cite{contractfuzzer} is the first black-box fuzzer for Ethereum smart contracts.
\textsc{ConFuzzius}~\cite{ConFuzzius} leverages constraint solving to generate inputs that satisfy complex conditions and generates meaningful sequences of inputs at runtime using dynamic data dependency analysis.
Most of these existing approaches are coverage-guided fuzzers that may waste testing resources on benign code areas and lack guidance in exploring the contract's state space.
By contrast, \textsc{Vulseye} is the first directed graybox fuzzer for smart contracts that concentrates testing resources on vulnerability-prone areas and performs directed fuzzing in both contract code and state spaces.

\textbf{Smart Contracts Analysis.}
\textsc{Vulseye} is related to work on smart contracts static analysis.
Numerous static analysis frameworks for smart contracts have been proposed~\cite{grech2018madmax,kalra2018zeus,DBLP:conf/ccs/TsankovDDGBV18,feist2019slither,GPTScan}.
\textsc{Zeus}~\cite{kalra2018zeus} is a symbolic model checking framework for verification of smart contract correctness and fairness policies.
\textsc{Securify}~\cite{DBLP:conf/ccs/TsankovDDGBV18} analyzes smart contracts using a dependency graph to check compliance and violation patterns that capture sufficient conditions for proving if a property holds or not.
\textsc{Slither}~\cite{feist2019slither} is a widely used tool that integrates IR transfer and flow analysis.
In our work, we use SlithIR as the foundation of static analysis, to identify code targets prior to conducting fuzz testing.

\subsection{Directed-Graybox Fuzzing Techniques}
\textsc{Vulseye} is closely related to work on directed-graybox techniques~\cite{aflgo,Hawkeye,10.1145/3377811.3380388,BEACON}.
\textsc{AFLGo}\cite{aflgo} is one of the first directed-graybox systems.
It utilizes a simulated annealing-based power schedule to drive the seed toward target sites.
\textsc{Hawkeye}~\cite{Hawkeye} boosts the speed for fuzzer to the target sites by utilizing power scheduling, adaptive mutation, and seed prioritization.
Existing directed-graybox fuzzers cannot be used to detect vulnerabilities in smart contracts.
Firstly, these tools lack the capability to autonomously identify vulnerable areas within smart contracts, relying instead on manual or external information to specify testing targets.
Additionally, their exclusive focus on the code space makes them less effective in testing stateful programs such as smart contracts.
\textsc{Vulseye} addresses these challenges and achieves effective stateful directed fuzzing.

\section{Conclusion}
\label{Conclusion}
In this paper, we propose \textsc{Vulseye}, a stateful directed fuzzer for smart contracts.
The key idea is to enhance fuzzing efficiency by exploring critical contract states through feedback from the contract state space and prioritizing testing resources to vulnerable code areas. 
To achieve this, we introduce code targets and state targets to flag the vulnerable contract code areas and states.
We design a novel fitness metric algorithm to steer the fuzzing toward these identified targets.
Experimental results show that \textsc{Vulseye} outperforms existing fuzzers and is effective in finding bugs in real-world scenarios.

\bibliographystyle{IEEEtran}
\bibliography{main}

\begin{thebibliography}{10}
\providecommand{\url}[1]{#1}
\csname url@samestyle\endcsname
\providecommand{\newblock}{\relax}
\providecommand{\bibinfo}[2]{#2}
\providecommand{\BIBentrySTDinterwordspacing}{\spaceskip=0pt\relax}
\providecommand{\BIBentryALTinterwordstretchfactor}{4}
\providecommand{\BIBentryALTinterwordspacing}{\spaceskip=\fontdimen2\font plus
\BIBentryALTinterwordstretchfactor\fontdimen3\font minus
  \fontdimen4\font\relax}
\providecommand{\BIBforeignlanguage}[2]{{%
\expandafter\ifx\csname l@#1\endcsname\relax
\typeout{** WARNING: IEEEtran.bst: No hyphenation pattern has been}%
\typeout{** loaded for the language `#1'. Using the pattern for}%
\typeout{** the default language instead.}%
\else
\language=\csname l@#1\endcsname
\fi
#2}}
\providecommand{\BIBdecl}{\relax}
\BIBdecl

\bibitem{bugs}
Z.~Zhang, B.~Zhang, W.~Xu, and Z.~Lin, ``Demystifying exploitable bugs in smart
  contracts,'' in \emph{Proceedings of International Conference on Software
  Engineering (ICSE)}, 2023.

\bibitem{DAO}
\BIBentryALTinterwordspacing
C.~Staff, ``What was the dao?'' 2023. [Online]. Available:
  \url{https://www.gemini.com/cryptopedia/the-dao-hack-makerdao}
\BIBentrySTDinterwordspacing

\bibitem{polynetwork}
\BIBentryALTinterwordspacing
REKT, ``Poly network - rekt,'' 2021. [Online]. Available:
  \url{https://www.gemini.com/cryptopedia/the-dao-hack-makerdao}
\BIBentrySTDinterwordspacing

\bibitem{10.1145/3512345}
X.~Zhu, S.~Wen, S.~Camtepe, and Y.~Xiang, ``Fuzzing: A survey for roadmap,''
  \emph{ACM Comput. Surv.}, vol.~54, 2022.

\bibitem{10548588}
S.~Wu, Z.~Li, L.~Yan, W.~Chen, M.~Jiang, C.~Wang, X.~Luo, and H.~Zhou, ``Are we
  there yet? unraveling the state-of-the-art smart contract fuzzers,'' in
  \emph{Proceedings of International Conference on Software Engineering
  (ICSE)}, 2024.

\bibitem{ILF}
J.~He, M.~Balunovi\'{c}, N.~Ambroladze, P.~Tsankov, and M.~Vechev, ``Learning
  to fuzz from symbolic execution with application to smart contracts,'' in
  \emph{Proceedings of Conference on Computer and Communications Security
  (CCS)}, 2019.

\bibitem{ECHIDNA}
G.~Grieco, W.~Song, A.~Cygan, J.~Feist, and A.~Groce, ``Echidna: Effective,
  usable, and fast fuzzing for smart contracts,'' in \emph{Proceedings of
  International Symposium on Software Testing and Analysis (ISSTA)}, 2020.

\bibitem{Harvey}
V.~W\"{u}stholz and M.~Christakis, ``Harvey: A greybox fuzzer for smart
  contracts,'' in \emph{Proceedings of Joint Meeting on European Software
  Engineering Conference and Symposium on the Foundations of Software
  Engineering (ESEC/FSE)}, 2020.

\bibitem{Targeted}
V.~Wustholz and M.~Christakis, ``Targeted greybox fuzzing with static lookahead
  analysis,'' in \emph{Proceedings of International Conference on Software
  Engineering (ICSE)}, 2020.

\bibitem{sFuzz}
T.~D. Nguyen, L.~H. Pham, J.~Sun, Y.~Lin, and Q.~T. Minh, ``Sfuzz: An efficient
  adaptive fuzzer for solidity smart contracts,'' in \emph{Proceedings of
  International Conference on Software Engineering (ICSE)}, 2020.

\bibitem{ConFuzzius}
C.~F. Torres, A.~K. Iannillo, A.~Gervais, and R.~State, ``Confuzzius: A data
  dependency-aware hybrid fuzzer for smart contracts,'' in \emph{Proceedings of
  European Symposium on Security and Privacy (EuroS\&P)}, 2021.

\bibitem{Smartian}
J.~Choi, D.~Kim, S.~Kim, G.~Grieco, A.~Groce, and S.~K. Cha, ``Smartian:
  Enhancing smart contract fuzzing with static and dynamic data-flow
  analyses,'' in \emph{Proceedings of International Conference on Automated
  Software Engineering (ASE)}, 2021.

\bibitem{Effectively}
J.~Su, H.-N. Dai, L.~Zhao, Z.~Zheng, and X.~Luo, ``Effectively generating
  vulnerable transaction sequences in smart contracts with reinforcement
  learning-guided fuzzing,'' in \emph{Proceedings of International Conference
  on Automated Software Engineering (ASE)}, 2023.

\bibitem{Rethinking}
Z.~Liu, P.~Qian, J.~Yang, L.~Liu, X.~Xu, Q.~He, and X.~Zhang, ``Rethinking
  smart contract fuzzing: Fuzzing with invocation ordering and important branch
  revisiting,'' \emph{IEEE Transactions on Information Forensics and Security
  (TIFS)}, vol.~18, 2023.

\bibitem{SmarTest}
S.~So, S.~Hong, and H.~Oh, ``{SmarTest}: Effectively hunting vulnerable
  transaction sequences in smart contracts through language {Model-Guided}
  symbolic execution,'' in \emph{Proceedings of USENIX Security Symposium
  (USENIX Security)}, 2021.

\bibitem{ITYFuzz}
C.~Shou, S.~Tan, and K.~Sen, ``Ityfuzz: Snapshot-based fuzzer for smart
  contract,'' in \emph{Proceedings of International Symposium on Software
  Testing and Analysis (ISSTA)}, 2023.

\bibitem{10.1145/3510003.3510230}
M.~B\"{o}hme, L.~Szekeres, and J.~Metzman, ``On the reliability of
  coverage-based fuzzer benchmarking,'' in \emph{Proceedings of International
  Conference on Software Engineering (ICSE)}, 2022.

\bibitem{sok}
\BIBentryALTinterwordspacing
P.~Wang and X.~Zhou, ``Sok: The progress, challenges, and perspectives of
  directed greybox fuzzing,'' \emph{CoRR}, vol. abs/2005.11907, 2020. [Online].
  Available: \url{https://arxiv.org/abs/2005.11907}
\BIBentrySTDinterwordspacing

\bibitem{10.1145/3133956.3134020}
M.~B\"{o}hme, V.-T. Pham, M.-D. Nguyen, and A.~Roychoudhury, ``Directed greybox
  fuzzing,'' in \emph{Proceedings of Conference on Computer and Communications
  Security (CCS)}, 2017.

\bibitem{aflgo}
------, ``Directed greybox fuzzing,'' in \emph{Proceedings of Conference on
  Computer and Communications Security (CCS)}, 2017.

\bibitem{Hawkeye}
H.~Chen, Y.~Xue, Y.~Li, B.~Chen, X.~Xie, X.~Wu, and Y.~Liu, ``Hawkeye: Towards
  a desired directed grey-box fuzzer,'' in \emph{Proceedings of Conference on
  Computer and Communications Security (CCS)}, 2018.

\bibitem{FISHFUZZ}
H.~Zheng, J.~Zhang, Y.~Huang, Z.~Ren, H.~Wang, C.~Cao, Y.~Zhang, F.~Toffalini,
  and M.~Payer, ``{FISHFUZZ}: Catch deeper bugs by throwing larger nets,'' in
  \emph{Proceedings of USENIX Security Symposium (USENIX Security)}, 2023.

\bibitem{SelectFuzz}
C.~Luo, W.~Meng, and P.~Li, ``Selectfuzz: Efficient directed fuzzing with
  selective path exploration,'' in \emph{Proceedings of {IEEE} Symposium on
  Security and Privacy (S\&P)}, 2023.

\bibitem{contractfuzzer}
B.~Jiang, Y.~Liu, and W.~K. Chan, ``Contractfuzzer: Fuzzing smart contracts for
  vulnerability detection,'' in \emph{Proceedings of International Conference
  on Automated Software Engineering (ASE)}, 2018.

\bibitem{clack2016smart}
C.~D. Clack, V.~A. Bakshi, and L.~Braine, ``Smart contract templates:
  foundations, design landscape and research directions,'' \emph{arXiv preprint
  arXiv:1608.00771}, 2016.

\bibitem{Ethereum}
\BIBentryALTinterwordspacing
G.~Wood, ``Ethereum: A secure decentralised generalised transaction ledger
  berlin version,'' 2022. [Online]. Available:
  \url{https://ethereum.github.io/yellowpaper/paper.pdf}
\BIBentrySTDinterwordspacing

\bibitem{sailfish}
P.~Bose, D.~Das, Y.~Chen, Y.~Feng, C.~Kruegel, and G.~Vigna, ``Sailfish:
  Vetting smart contract state-inconsistency bugs in seconds,'' in
  \emph{Proceedings of {IEEE} Symposium on Security and Privacy (S\&P)}, 2022.

\bibitem{godefroid2008automated}
P.~Godefroid, M.~Y. Levin, D.~A. Molnar \emph{et~al.}, ``Automated whitebox
  fuzz testing.'' in \emph{Proceedings of Network and Distributed System
  Security Symposium (NDSS)}, 2008.

\bibitem{DBLP:conf/uss/KimKBBKT23}
K.~Kim, S.~Kim, K.~R.~B. Butler, A.~Bianchi, R.~Kennell, and D.~Tian, ``Fuzz
  the power: Dual-role state guided black-box fuzzing for {USB} power
  delivery,'' in \emph{Proceedings of USENIX Security Symposium (USENIX
  Security)}, 2023.

\bibitem{8371326}
H.~Liang, X.~Pei, X.~Jia, W.~Shen, and J.~Zhang, ``Fuzzing: State of the art,''
  \emph{IEEE Transactions on Reliability}, vol.~67, 2018.

\bibitem{281370}
J.~Ba, M.~B{\"o}hme, Z.~Mirzamomen, and A.~Roychoudhury, ``Stateful greybox
  fuzzing,'' in \emph{Proceedings of USENIX Security Symposium (USENIX
  Security)}, 2022.

\bibitem{feist2019slither}
J.~Feist, G.~Grieco, and A.~Groce, ``Slither: a static analysis framework for
  smart contracts,'' in \emph{IEEE/ACM International Workshop on Emerging
  Trends in Software Engineering for Blockchain (WETSEB)}, 2019.

\bibitem{8967437}
M.~Nachtigall, L.~Nguyen Quang~Do, and E.~Bodden, ``Explaining static analysis
  - a perspective,'' in \emph{Proceedings of International Conference on
  Automated Software Engineering Workshop (ASEW)}, 2019.

\bibitem{10.1145/2976749.2978309}
L.~Luu, D.-H. Chu, H.~Olickel, P.~Saxena, and A.~Hobor, ``Making smart
  contracts smarter,'' in \emph{Proceedings of Conference on Computer and
  Communications Security (CCS)}, 2016.

\bibitem{10.1145/3274694.3274737}
C.~F. Torres, J.~Sch{\"{u}}tte, and R.~State, ``Osiris: Hunting for integer
  bugs in ethereum smart contracts,'' in \emph{Proceedings of Annual Computer
  Security Applications Conference (ACSAC)}, 2018.

\bibitem{DBLP:conf/acsac/NikolicKSSH18}
I.~Nikolic, A.~Kolluri, I.~Sergey, P.~Saxena, and A.~Hobor, ``Finding the
  greedy, prodigal, and suicidal contracts at scale,'' in \emph{Proceedings of
  Annual Computer Security Applications Conference (ACSAC)}, 2018.

\bibitem{networkx}
\BIBentryALTinterwordspacing
NetworkX, ``networkx,'' 2023. [Online]. Available: \url{https://networkx.org/}
\BIBentrySTDinterwordspacing

\bibitem{GA}
V.~Chahar, S.~Katoch, and S.~Chauhan, ``A review on genetic algorithm: Past,
  present, and future,'' \emph{Multimedia Tools and Applications}, vol.~80,
  2021.

\bibitem{py-evm}
\BIBentryALTinterwordspacing
ethereum, ``py-evm,'' 2023. [Online]. Available:
  \url{https://github.com/ethereum/py-evm}
\BIBentrySTDinterwordspacing

\bibitem{xFuzz}
Y.~Xue, J.~Ye, W.~Zhang, J.~Sun, L.~Ma, H.~Wang, and J.~Zhao, ``xfuzz: Machine
  learning guided cross-contract fuzzing,'' \emph{IEEE Transactions on
  Dependable and Secure Computing (TDSC)}, 2022.

\bibitem{durieux2020empirical}
T.~Durieux, J.~F. Ferreira, R.~Abreu, and P.~Cruz, ``Empirical review of
  automated analysis tools on 47,587 ethereum smart contracts,'' in
  \emph{Proceedings of International conference on software engineering
  (ICSE)}, 2020.

\bibitem{AFL}
\BIBentryALTinterwordspacing
google, ``american fuzzy lop,'' 2022. [Online]. Available:
  \url{https://github.com/google/AFL}
\BIBentrySTDinterwordspacing

\bibitem{swc}
\BIBentryALTinterwordspacing
SmartContractSecurity, ``Smart contract weakness classification,'' 2020.
  [Online]. Available: \url{https://swcregistry.io/}
\BIBentrySTDinterwordspacing

\bibitem{Etherscan}
\BIBentryALTinterwordspacing
Etherscan, ``Etherscan.io,'' 2023. [Online]. Available:
  \url{https://etherscan.io/}
\BIBentrySTDinterwordspacing

\bibitem{bitdegree}
\BIBentryALTinterwordspacing
BitDegree, ``bitdegree,'' 2023. [Online]. Available:
  \url{https://cn.bitdegree.org/}
\BIBentrySTDinterwordspacing

\bibitem{onchainFuzz}
M.~Ye, Y.~Nan, Z.~Zheng, D.~Wu, and H.~Li, ``Detecting state inconsistency bugs
  in dapps via on-chain transaction replay and fuzzing,'' in \emph{Proceedings
  of International Symposium on Software Testing and Analysis (ISSTA)}, 2023.

\bibitem{grech2018madmax}
N.~Grech, M.~Kong, A.~Jurisevic, L.~Brent, B.~Scholz, and Y.~Smaragdakis,
  ``Madmax: Surviving out-of-gas conditions in ethereum smart contracts,''
  \emph{Proceedings of the ACM on Programming Languages}, vol.~2, no. OOPSLA,
  2018.

\bibitem{kalra2018zeus}
S.~Kalra, S.~Goel, M.~Dhawan, and S.~Sharma, ``{ZEUS:} analyzing safety of
  smart contracts,'' in \emph{Proceedings of Network and Distributed System
  Security Symposium (NDSS)}, 2018.

\bibitem{DBLP:conf/ccs/TsankovDDGBV18}
P.~Tsankov, A.~M. Dan, D.~Drachsler{-}Cohen, A.~Gervais, F.~B{\"{u}}nzli, and
  M.~T. Vechev, ``Securify: Practical security analysis of smart contracts,''
  in \emph{Proceedings of Conference on Computer and Communications Security
  (CCS)}, 2018.

\bibitem{GPTScan}
Y.~Sun, D.~Wu, Y.~Xue, H.~Liu, H.~Wang, Z.~Xu, X.~Xie, and Y.~Liu, ``Gptscan:
  Detecting logic vulnerabilities in smart contracts by combining gpt with
  program analysis,'' in \emph{Proceedings of International Conference on
  Software Engineering (ICSE)}, 2024.

\bibitem{10.1145/3377811.3380388}
V.~W\"{u}stholz and M.~Christakis, ``Targeted greybox fuzzing with static
  lookahead analysis,'' in \emph{Proceedings of International Conference on
  Software Engineering (ICSE)}, 2020.

\bibitem{BEACON}
H.~Huang, Y.~Guo, Q.~Shi, P.~Yao, R.~Wu, and C.~Zhang, ``Beacon: Directed
  grey-box fuzzing with provable path pruning,'' in \emph{Proceedings of {IEEE}
  Symposium on Security and Privacy (S\&P)}, 2022.

\end{thebibliography}

\begin{IEEEbiography}
  [{\includegraphics [width=1in,height=1.25in, clip,keepaspectratio]{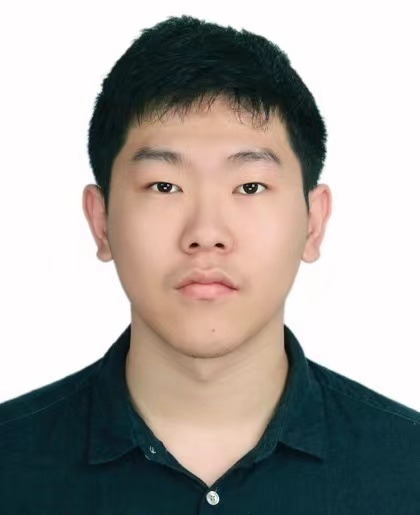}}]{Ruichao Liang}
  received the B.E. degree in cyberspace security from Wuhan University, Wuhan, China, in 2020.
  He is currently working toward the Ph.D. degree with the School of Cyber Science and Engineering, Wuhan University, China.
  His research interests include web3 security and vulnerability analysis.

\end{IEEEbiography}

\vspace{-30pt}

\begin{IEEEbiography}[{\includegraphics[width=1in,height=1.25in,clip,keepaspectratio]{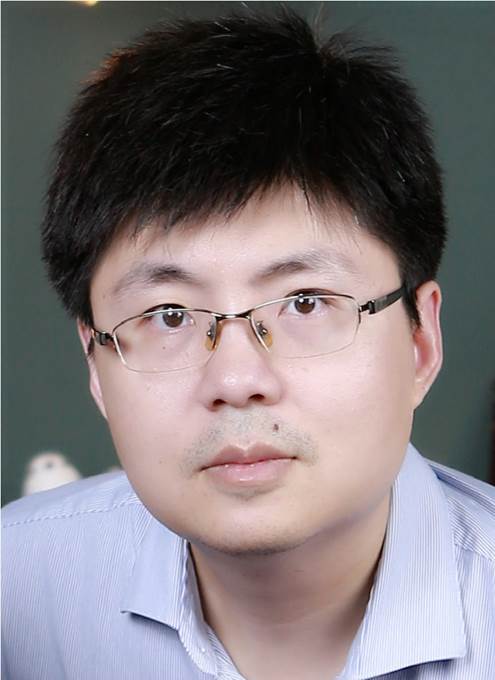}}]{Jing Chen}
  
  received the Ph.D degree in computer science from the Huazhong University of Science and Technology, Wuhan. 
  He is currently a full professor with the School of Cyber Science and Engineering, Wuhan University. 
  He is a senior member of IEEE and served as the vice chair of the ACM Turing Award Celebration Conference (TURC) 2023. 
  He has published more than 150 research papers in many international journals and conferences, including USENIX Security, ACM CCS, INFOCOM, IEEE TDSC, IEEE TIFS, IEEE TMC, IEEE TC, IEEE TPDS, IEEE TSC, etc. 
  He was twice runner-up for the best paper at INFOCOM 2018 and INFOCOM 2021. 
  His research interests include the areas of network security, cloud security, and mobile security.
\end{IEEEbiography}

\vspace{-30pt}

\begin{IEEEbiography}[{\includegraphics[width=1in,height=1.25in,clip,keepaspectratio]{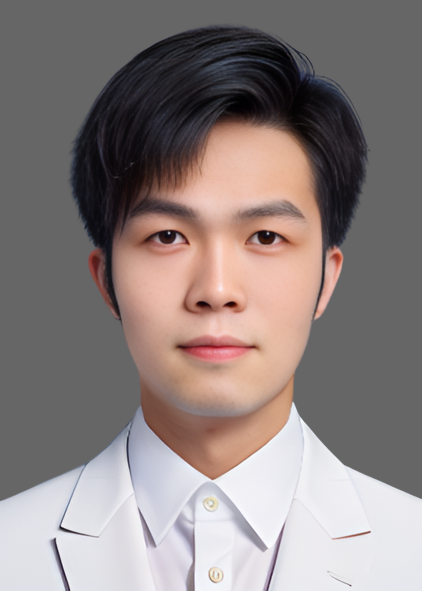}}]{Cong Wu}
  received the Ph.D. degree from the School of Cyber Science and Engineering, Wuhan University, in 2022. 
  He is currently a Research Fellow with the Cyber Security Laboratory, Nanyang Technological University, Singapore. 
  His leading research outcomes have appeared in USENIX Security, ACM CCS, IEEE TRANSACTIONS ON DEPENDABLE AND SECURE COMPUTING, and IEEE TRANSACTIONS ON MOBILE COMPUTING. 
  His research interests include biometric security, system security, mobile security, and web3 security.

\end{IEEEbiography}
\vspace{-30pt}

\begin{IEEEbiography}[{\includegraphics[width=1in,height=1.25in,clip,keepaspectratio]{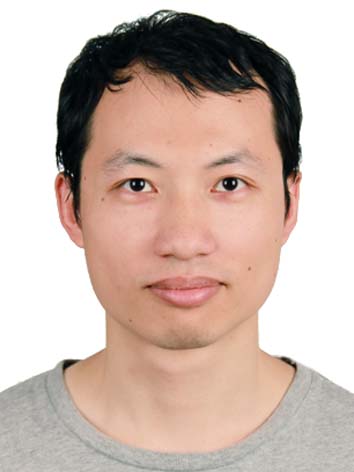}}]{Kun He}
  received his Ph.D. from Wuhan University, Wuhan, China. 
  He is currently an associate professor with Wuhan University. 
  His research interests include cryptography and data security. 
  He has published more than 30 research papers in various journals and conferences, such as TIFS, TDSC, TMC, USENIX Security, CCS, and INFOCOM.
  
\end{IEEEbiography}

\vspace{-30pt}

\begin{IEEEbiography}
  [{\includegraphics [width=1in,height=1.25in, clip,keepaspectratio]{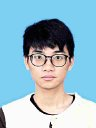}}]{Yueming Wu}
  received the B.E. degree in Computer Science and Technology at Southwest Jiaotong University, Chengdu, China, in 2016 and the Ph.D. degree in School of Cyber Science and Engineering at Huazhong University of Science and Technology, Wuhan, China, in 2021.
  He is currently a research fellow in the School of Computer Science and Engineering at Nanyang Technological University. 
  His primary research interests lie in malware analysis and vulnerability analysis.
\end{IEEEbiography}
\vspace{-30pt}

\begin{IEEEbiography}
  [{\includegraphics [width=1in,height=1.25in, clip,keepaspectratio]{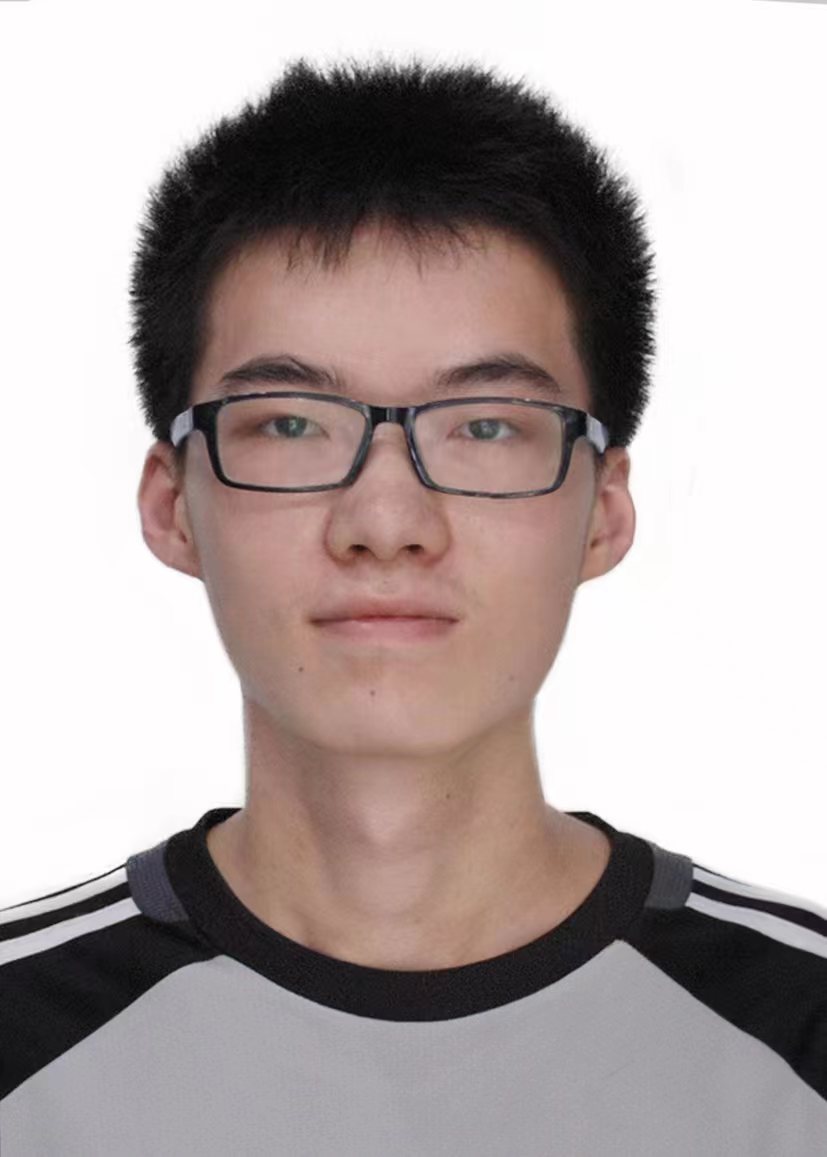}}]{Ruochen Cao}
  received the B.E. degree in cyberspace security from Wuhan University, Wuhan, China, in 2023.
  He is currently working toward the master's degree with the School of Cyber Science and Engineering, Wuhan University, China.

\end{IEEEbiography}
\vspace{-90pt}

\begin{IEEEbiography}[{\includegraphics[width=1in,height=1.25in,clip,keepaspectratio]{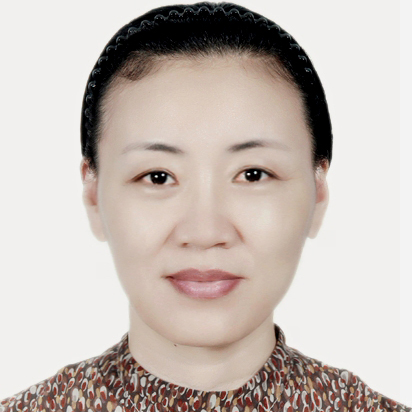}}]{Ruiying Du}
  received the BS, MS, PH. D degrees in computer science in 1987, 1994 and 2008, from Wuhan University, Wuhan, China. 
  She is a professor at School of Cyber Science and Engineering, Wuhan University. 
  Her research interests include network security, wireless network, cloud computing and mobile computing. 
  She has published more than 80 research papers in many international journals and conferences, such as TPDS, USENIX Security, CCS, INFOCOM, SECON, TrustCom, NSS.
\end{IEEEbiography}

\vspace{-100pt}

\begin{IEEEbiography}
  [{\includegraphics [width=1in,height=1.25in, clip,keepaspectratio]{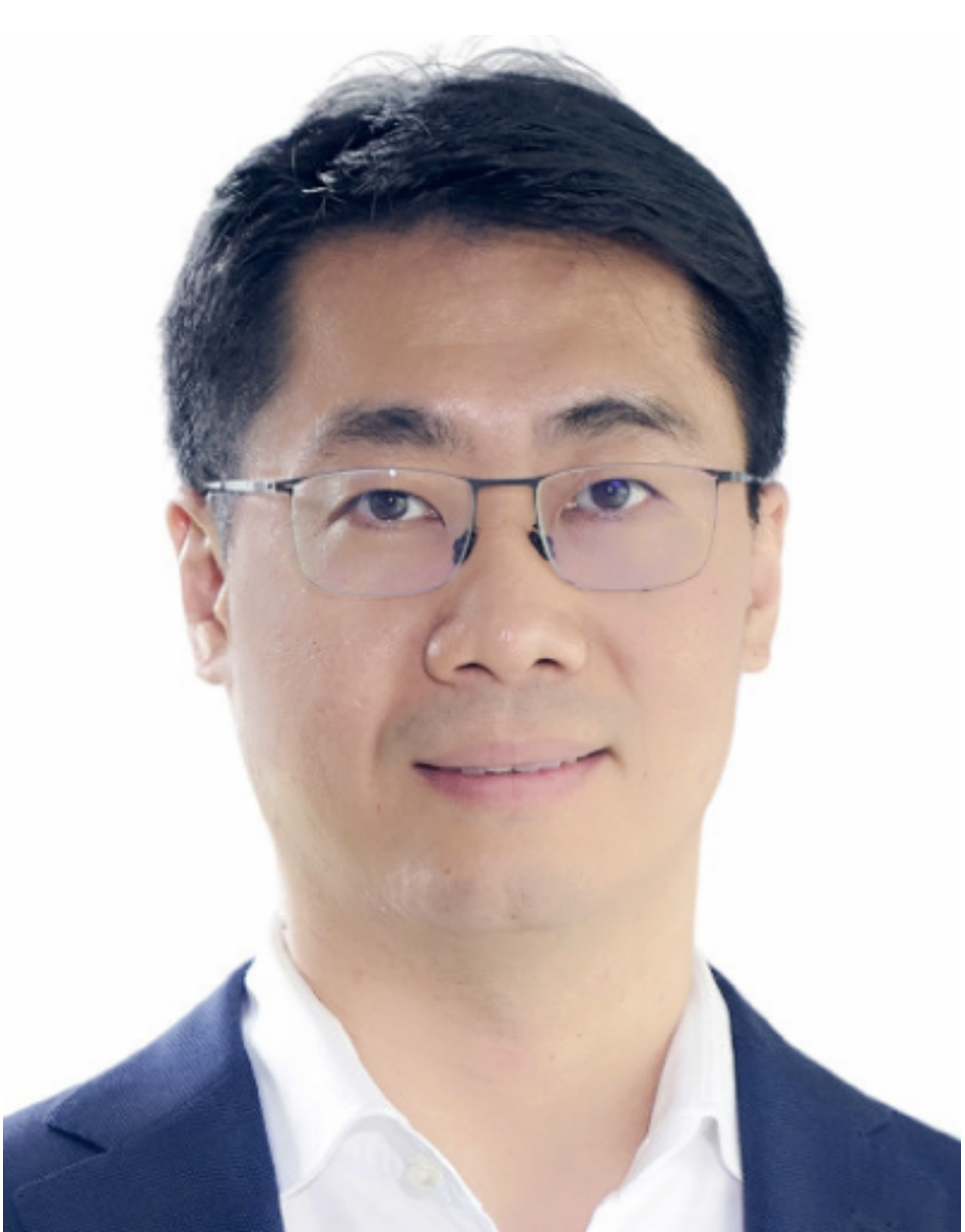}}]{Yang Liu}
  (Senior Member, IEEE) is currently a Full Professor and the Director of the Cyber Security Laboratory, Nanyang Technological University, Singapore. 
  He specializes in software security, verification, software engineering, and artificial intelligence. His research has bridged the gap between the theory and practical usage of formal methods and program analysis to evaluate the design and implementation of software for high assurance and security. His work led to the development of state-of-the-art model checker and process analysis toolkit (PAT). 
  He has more than 200 publications and six best paper awards in top tier conferences and journals. 
  With more than 50 million Singapore dollar funding support, he is leading a large research team working on the state-of-the-art software engineering and cyber security problems.

\end{IEEEbiography}
\vspace{-95pt}

\begin{IEEEbiography}
  [{\includegraphics [width=1in,height=1.25in, clip,keepaspectratio]{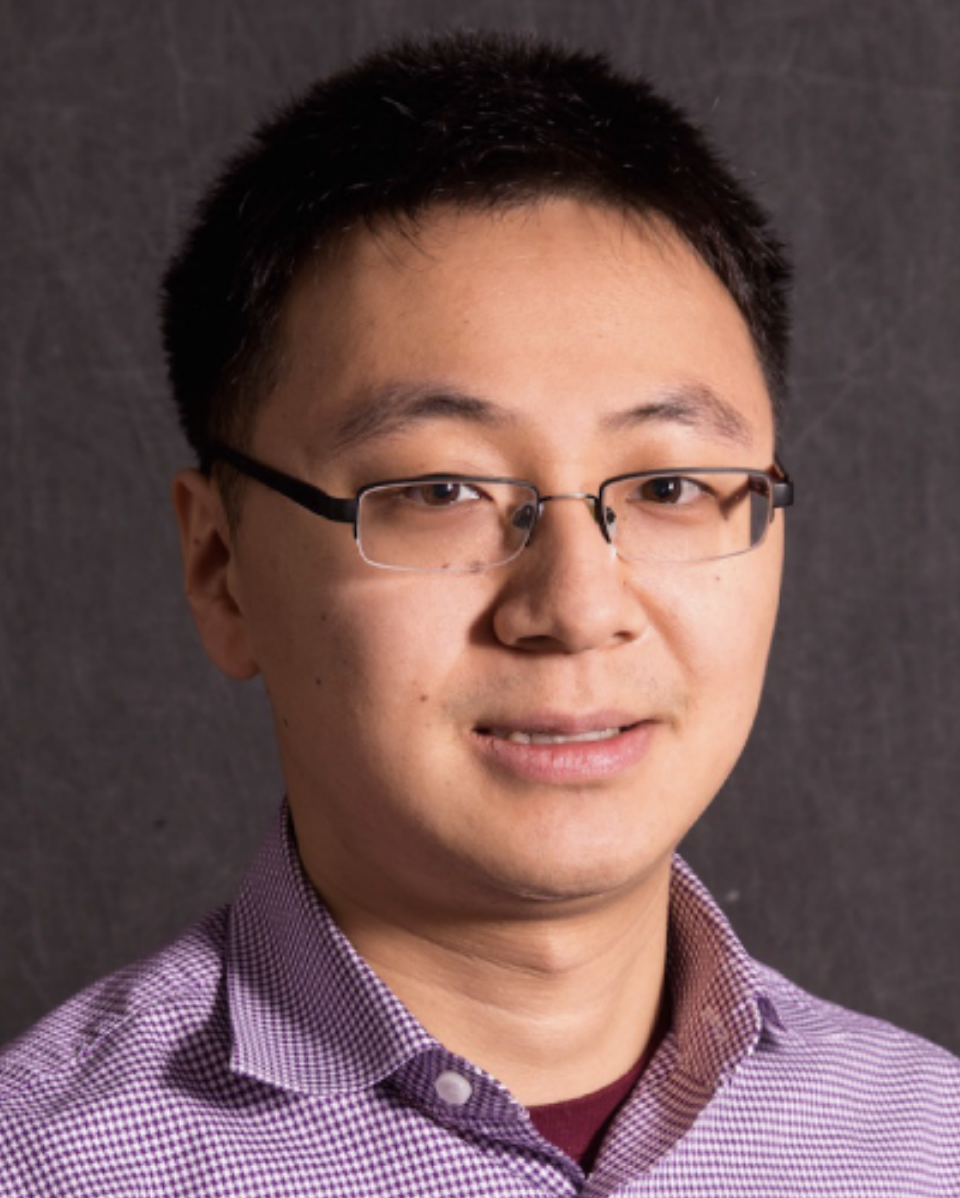}}]{Ziming Zhao}
  received the Ph.D. degree in computer science from Arizona State University, Tempe, AZ, USA, in 2014. He is an Associate Professor with the Khoury College of Computer Sciences and the Director of the CyberspACe securiTy and forensIcs Laboratory (CactiLab), Northeastern University, Boston, USA. 
  His research has been supported by the U.S. National Science Foundation (NSF), the U.S. Department of Defense, the U.S. Air Force Office of Scientific Research, and the U.S. National Centers of Academic Excellence in Cybersecurity (part of the National Security Agency). 
  His research outcomes have appeared in IEEE SECURITY AND PRIVACY, USENIX Security, ACM CCS, NDSS, ACM MobiSys, ACM/IEEE DAC, IEEE RTAS, ACM TISSEC/TOPS, IEEE TRANSACTIONS ON DEPENDABLE AND SECURE COMPUTING, and IEEE TRANSACTIONS ON INFORMATION FORENSICS AND SECURITY. 
  His current research interests include systems and software security, network and web security, and human-centric security. 
  He was a recipient of the NSF CAREER Award and the NSF CRII Award. 
  He was also a recipient of the Test-ofTime Paper Award at ACM SACMAT 2024. 
  Additionally, he has received Best/Distinguished Paper Awards from several prestigious conferences, including USENIX Security 2019, ACM AsiaCCS 2022, ACM CODASPY 2014, and ITU Kaleidoscope 2016.
\end{IEEEbiography}

\end{document}